\documentclass[sw]{iosart2x}

\pubyear{0000}
\volume{0}
\firstpage{1}
\lastpage{1}

\usepackage{diagbox}
\usepackage{float}
\usepackage{adjustbox}

\usepackage[usenames,dvipsnames]{xcolor}
\usepackage{caption}
\usepackage{listings}
\usepackage{colortbl}
\usepackage{paralist}
\usepackage{enumitem,kantlipsum}
\usepackage{wrapfig}
\usepackage{tabularx}
\usepackage{multirow,array}
\usepackage{todonotes}
\usepackage{amsmath}
\usepackage{subcaption}
\usepackage{graphicx}
\usepackage{xspace}
\usepackage{nccmath}

\usepackage{algorithmicx}
\usepackage{algorithm}
\usepackage[noend]{algpseudocode}

\algnewcommand{\algorithmicand}{\textbf{ and }}
\algnewcommand{\algorithmicor}{\textbf{ or }}
\algnewcommand{\Or}{\algorithmicor}
\algnewcommand{\And}{\algorithmicand}
\algnewcommand{\var}{\texttt}

\usepackage{graphicx}
\newcommand{\pap}{PAPyA\xspace}

\usepackage[dvipsnames]{xcolor}
\usepackage{listings}

\newcommand\YAMLcolonstyle{\color{red}\mdseries}
\newcommand\YAMLkeystyle{\color{black}\bfseries}
\newcommand\YAMLvaluestyle{\color{blue}\mdseries}

\makeatletter

\newcommand\language@yaml{yaml}

\expandafter\expandafter\expandafter\lstdefinelanguage
\expandafter{\language@yaml}
{
  keywords={true,false,null,y,n},
  keywordstyle=\color{darkgray}\bfseries,
  basicstyle=\YAMLkeystyle,                                 
  sensitive=false,
  comment=[l]{\#},
  morecomment=[s]{/*}{*/},
  commentstyle=\color{purple}\ttfamily,
  stringstyle=\YAMLvaluestyle\ttfamily,
  moredelim=[l][\color{orange}]{\&},
  moredelim=[l][\color{magenta}]{*},
  moredelim=**[il][\YAMLcolonstyle{:}\YAMLvaluestyle]{:},   
  morestring=[b]',
  morestring=[b]",
  literate =    {---}{{\ProcessThreeDashes}}3
                {>}{{\textcolor{red}\textgreater}}1     
                {|}{{\textcolor{red}\textbar}}1 
                {\ -\ }{{\mdseries\ -\ }}3,
}

\lst@AddToHook{EveryLine}{\ifx\lst@language\language@yaml\YAMLkeystyle\fi}
\makeatother

\newcommand\ProcessThreeDashes{\llap{\color{cyan}\mdseries-{-}-}}

 \definecolor{backcolour}{rgb}{0.95,0.95,0.92}

\lstdefinestyle{turtleStyle}{
    numberstyle=\tiny\color{lightgray},
    basicstyle=\scriptsize\ttfamily,
    breakatwhitespace=false,         
    breaklines=true,                 
    captionpos=b,                    
    keepspaces=false,                 
    numbers=right,                    
    numbersep=5pt,                  
    showspaces=false,                
    showstringspaces=false,
    showtabs=false,                  
  mathescape=true,
 columns=fixed,
 basewidth=0.5em,
 escapeinside={(*}{*)},
 showstringspaces=false,
 tabsize=2,
 breaklines=true,
 frame=tbrl,
 frameround=ffff,
 backgroundcolor={\color{backcolour}},
 rulecolor={\color{lightgray}},
 captionpos=b,
 xleftmargin=0em,
 framexleftmargin=0em
}

\newtheorem{definition}{Definition}

\begin{document}

\begin{frontmatter}

\title{\texttt{PAPyA}: a Library for Performance Analysis of SQL-based RDF Processing Systems}
\runtitle{PAPyA: Performance Analysis of Large RDF Graphs Processing Made Easy}


\begin{aug}
\author[A]{\inits{M.}\fnms{Mohamed} \snm{Ragab}\ead[label=e1]{mohamed.ragab@ut.ee}%
}
\author[B]{\inits{A.S.}\fnms{Adam Satria} \snm{Adidarma}\ead[label=e2]{adam.19051@mhs.its.ac.id}
\thanks{Equal Contribution with the first author. Corresponding Author mail \printead{e1}. }.}
\author[C]{\inits{R.-T.}\fnms{Riccardo} \snm{Tommasini}\ead[label=e3]{riccardo.tommasini@ut.ee}}

\address[A]{Computer Science Inst., \orgname{University of Tartu},
Tartu, \cny{Estonia}\printead[presep={\\}]{e1}}
\address[B]{Sepuluh Nopember Institute of Technology,
Surabaya, \cny{Indonesia}\printead[presep={\\}]{e2}}
\address[C]{LIRIS Lab \orgname{INSA de Lyon},
Villeurbanne, \cny{France}\printead[presep={\\}]{e3}}

\end{aug}


\begin{abstract}
\textit{Prescriptive Performance Analysis} (PPA) has shown to be more useful than traditional \textit{descriptive} and \textit{diagnostic} analyses for making sense of Big Data (BD) frameworks' performance. In practice, when processing large (RDF) graphs on top of relational BD systems, several design decisions emerge and cannot be decided automatically, e.g., the choice of the schema, the partitioning technique, and the storage formats. PPA, and in particular ranking functions, helps enable actionable insights on performance data, leading practitioners to an easier choice of the best way to deploy BD frameworks, especially for graph processing. However, the amount of experimental work required to implement PPA is still huge. In this paper, we present \pap~\footnote{\url{https://github.com/DataSystemsGroupUT/PAPyA}}, a library for implementing PPA that allows (1) preparing RDF graphs data for a processing pipeline over relational BD systems, (2) enables automatic ranking of the performance in a \textit{user-defined} solution space of experimental dimensions; (3) allows user-defined flexible extensions in terms of systems to test and ranking methods. We showcase \pap on a set of experiments based on the SparkSQL framework. \pap simplifies the performance analytics of BD systems for processing large (RDF) graphs. We provide \pap~as a public \textit{open-source} library under an~\textit{MIT} license that will be a catalyst for designing new research prescriptive analytical techniques for BD applications.

\end{abstract}


\end{frontmatter}


\section{Introduction}

The increasing adoption of Knowledge Graphs (KGs) in industry and academia requires scalable systems for taming linked data at large volumes and velocity. In absence of a scalable native graph system for \textit{querying} large (RDF) graphs~\cite{cacm21}, most approaches fall back to using relational Big Data (BD) frameworks (e.g., \textit{Apache Spark} or \textit{Impala}) for handling large graph query workloads~\cite{schatzle2016s2rdf,schatzle2014sempala}. Despite its flexibility, the relational model requires several additional \textit{design decisions} when used for processing graphs, which cannot be decided automatically, e.g., the choice of the \textit{schema}, the \textit{partitioning techniques}, and the \textit{storage formats}.

In~\cite{ragab2021depth}, we highlight the severity of the problem by showing the lack of performance replicability of BD systems for querying large (RDF) graphs. In particular, we showed that changing even just one experimental parameter, e.g., \textit{partitioning technique} or \textit{storage encoding}, invalidates existing optimizations in the relational representation of RDF data. 
We observe that issues do not lay in the way the investigations were conducted but rather in the maturity of the analysis, which is limited to \textit{descriptive} or at most \textit{diagnostic} observations of the system behavior. Such discussions leave much work for practitioners to transform performance observations into actionable insights~\cite{ragabieee21}.

\begin{wrapfigure}[9]{r}{0.25\textwidth}
\vspace{-10pt}
    \centering
      \includegraphics[width=0.99\linewidth]{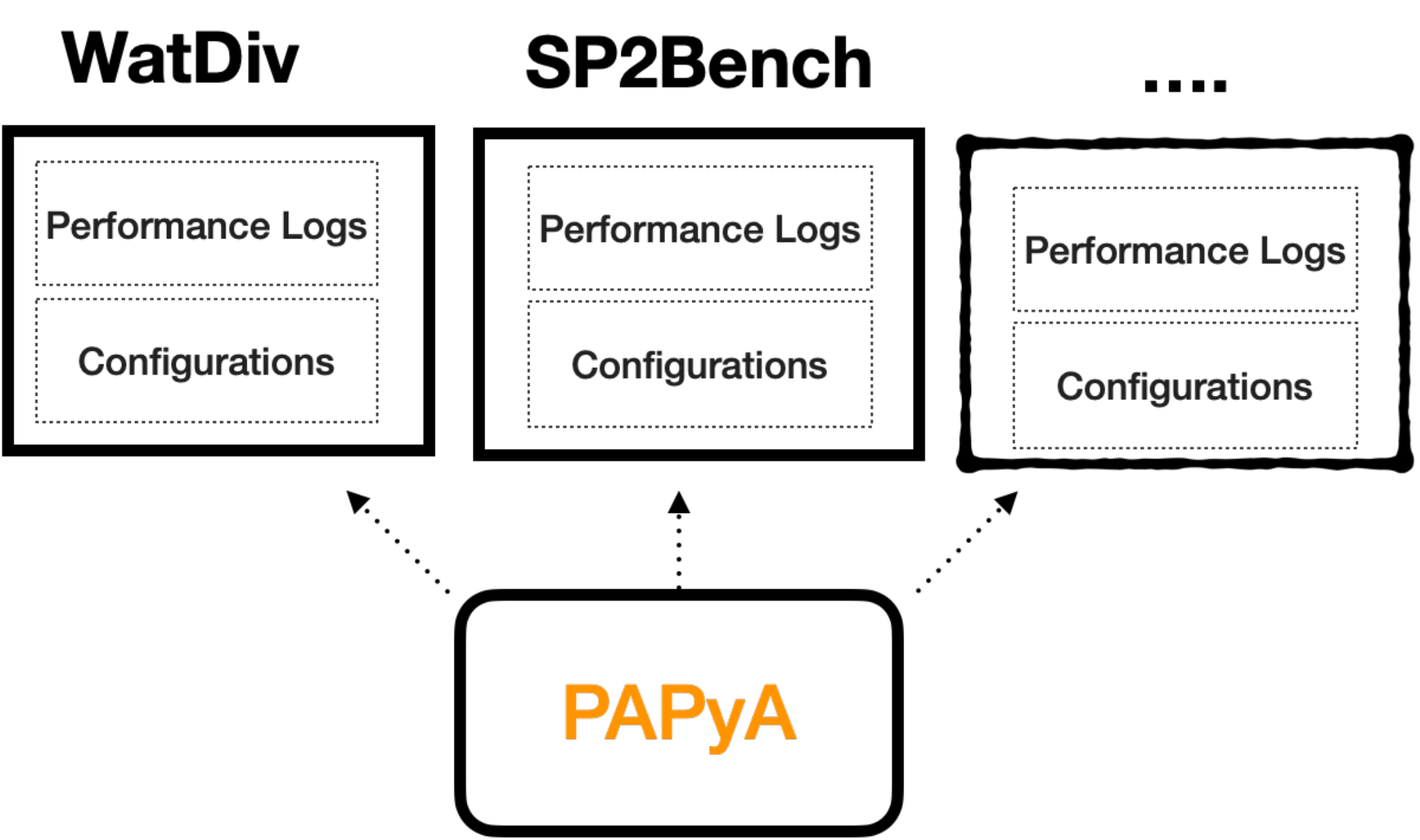}
      \caption{\pap dynamicity with RDF benchmarks.}
      \label{fig:papyadynamicity}
\end{wrapfigure} 

Later in~\cite{ragabieee21}, we have introduced the concept of \textit{Bench-Ranking} as a means for enabling \textit{Prescriptive Performance Analysis} (PPA) for processing larger RDF graphs. The PPA is an alternative to descriptive/diagnostic analyses that aims to answer the question \textit{What should we do?}~\cite{lepenioti2020prescriptive}. In practice, \textit{Bench-Ranking} enables informed decision-making without neglecting the effectiveness of the performance analyses~\cite{ragabieee21}. In particular, we showed how it could prescribe the best-performing combination of schema, partitioning technique, and storage format for querying large (RDF) graphs on top of SparkSQL framework~\cite{ragabieee21}.

Our direct experience with big RDF graphs processing shows that the most \textit{time-consuming} phases were the \textit{data preparation} and \textit{performance analytics}. Although the Bench-Ranking methodology simplifies performance analyses, its current implementation is still limited
as it does not follow any specific software engineering best practices. Thus, practitioners who implement the Bench-Ranking methodology may face the following challenges:

\begin{enumerate} [leftmargin=*, label=\textbf{C.\arabic*}]

\item \textbf{Experiment Preparation} requires huge data engineering efforts to build the full pipeline for processing large graphs on top of BD systems, to put the data in the \textit{logical} and \textit{physical} representations that adapt with \textit{relational distributed} environments. Moreover, the current experimental preparation in Bench-Ranking requires incorporating several systems, e.g., \textit{Apache Jena} (i.e., for logical schema definitions), and \textit{Apache SparkSQL} (i.e., for physical partitioning and storage).

\item \textbf{Portability and Usability}: deciding new requirements in the Bench-Ranking framework's current implementation (e.g., changes over the number of tasks (i.e., queries) or changes over the experimental dimensions/configurations.) would lead to repeating vast parts of the work.

\item \textbf{Flexibility and Extensibility}: the current implementation of the Bench-Ranking framework does not fully reflect the flexibility and extensibility of the framework in terms of experimental dimensions and ranking criteria extensibility.

\item  \textbf{Complexity and Compoundnesss}: practitioners may find Bench-Ranking criteria and evaluation metrics quite complex to implement. Moreover, the current implementation does not provide an interactive interface that eases interconnections of various modules of the framework (e.g., data and performance visualization).     
\end{enumerate}

%

To mitigate these problems, we design and implement an open-source library called \textit{\pap} (Prescriptive Analysis in Python Actions), which contributes with
(1) reducing the engineering work required for graph processing preparations and data loading; 
(2) reproducing existing experiments (according to user needs and convenience) for relational processing of SPARQL queries using SparkSQL. This will lead to reducing massive efforts for building analytical pipelines from scratch for relational BD systems that are subject to the experiments. Moreover, \pap also aims at
(3) automating the Bench-Ranking methods for prescriptive performance analysis described in~\cite{ragabieee21}. In practice, \pap facilitates navigating the complex solution space via packaging the functionality of different ranking functions as well as \textit{Multi-Optimization} (MO) techniques into \textit{interactive} programmatic library interfaces. Last but not least, 
(4) checking the replicability of the relational BD systems' performance for querying large (RDF) graphs within a complex experimental solution space.

The focus of this paper is to show the internals and functionality of \pap as a means for providing PPA for BD relational systems that query large (RDF) graphs. For completeness, we aim to describe \pap prescriptions with the WatDiv benchmark~\cite{alucc2014diversified} experiments~\footnote{Due to space limits, we diagnose the WatDiv prescriptions results and evaluation metrics in section~\ref{sec:bench_in_use} on the library GitHub page, mentioned below.}. In~\cite{ragabieee21}, we applied Bench-Ranking to the SP$^2$B benchmark~\cite{schmidt2009sp}. However, the best-performing configurations depend on the dataset and the query workload~\cite{ragabieee21}. Thanks to \pap, we can easily perform the PPA for any RDF benchmark, pointing out to the performance results, and specifying the experimental configurations (see Figure~\ref{fig:papyadynamicity}).    

\vspace{3pt}

\noindent\textit{\textbf{Outline.}}
Section~\ref{sec:bg} presents the necessary preliminaries to understand the paper's content.
Section~\ref{sec:benchranking} briefly introduces the \textit{Bench-Ranking} framework concepts~\cite{ragabieee21}.
Section~\ref{sec:lib_req_arch} presents the \pap requirements alongside its framework architecture.
Section~\ref{sec:bench_in_use} shows how to use \pap in practice, providing examples from existing experiments~\cite{ragab2021depth,ragabieee21}. 
Section~\ref{sec:relwork} discusses related work, while Section~\ref{sec:conclusion} concludes the paper and presents \pap's roadmap.

\section{Background}\label{sec:bg}

This section presents the necessary background to understand the paper's content. We assume that the reader is familiar with the RDF data model and the SPARQL query language.

\subsection{RDF Relational Processing Experimental Dimensions}
Several design decisions emerge when utilizing relational BD systems for querying large (RDF) graphs, such as the relational schema, partitioning techniques, and storage formats. These experimental dimensions directly impact the performance of BD systems while querying large graphs. Intuitively, these dimensions entail different choice options (we call them dimensions' parameters).

First, the \textbf{relational schema}, it is easy to show that we have different options on how to represent graphs as relational tables, and the choice hugely impacts the performance~\cite{abadi2007scalable,schatzle2016s2rdf}. We identify the most used ones in the literature of RDF processing~\cite{abdelaziz2017survey,abadi2007scalable,schatzle2016s2rdf,schatzle2014sempala}: \textit {Single Statement} Table (ST) schema that stores RDF triples in a \textit{ternary}-column relation (subject, predicate, object), and often requires many self-joins; \textit {Vertically Partitioned} Tables (VP) proposed by \textit{Abadi et.al.}~\cite{abadi2007scalable} to mitigate issues of \textit{self-joins} in ST schema proposing to use \textit{binary} relations (subject, object) for each unique predicate in the RDF dataset; the \textit {Property Tables} (PT) schema that prescribes clustering multiple RDF properties as \textit{n-ary} columns table for the same \textit{subject} to group entities that are similar in structure. Lastly, two more RDF relational schema advancements also emerge in the literature. The Wide Property Table (WPT) schema encodes the entire dataset into a single \textit{denormalized} table. WPT is initially proposed for \textit{Sempala} system by~\textit{Sc\"atzle et.al.}~\cite{schatzle2014sempala}, who also proposed another schema optimization that extended the version of the VP schema (ExtVP)~\cite{schatzle2016s2rdf} that pre-computes \textit{semi-join} VP tables to reduce data shuffling. 

BD platforms are designed to scale horizontally; thus, data \textit{partitioning} is another crucial dimension for querying large graphs. However, choosing the right \textbf{partitioning} technique for RDF data is non-trivial. To this extent, we followed the indication of \textit{Akhter et al.}~\cite{akhter2018empirical}. In particular, we selected the three techniques that can be applied directly to RDF graphs while being mapped to a relational schema. Namely, (i) \textit{Horizontal Partitioning} (HP) randomly divides data evenly on the number of machines in the cluster, i.e., $n$ equally sized chunks, where $n$ is the number of machines in the cluster. (ii) {Subject-Based Partitioning} (SBP) (or (iii) Predicate-Based Partitioning (PBP)) distributes triples to the various partitions according to the \textit{hash value} computed for the \textit{subjects} (\textit{predicates}). As a result, all the triples with the same \textit{subject} (\textit{predicate}) reside on the same partition. Notably, both SBP and PBP may suffer from data skewness which impacts parallelism.

Serializing RDF data also offers many options such as \textit{RDF/XML}, \textit{Turtle}, \textit{JSON-LD}, to name a few. On the same note, BD platforms offer many options for reading/writing to various file formats and storage backends. Therefore, we need to consider the variety of \textbf{storage} formats~\cite{ivanov2019impact}. We specifically focus on the various \textit{Hadoop Distributed File System }(HDFS) file formats that are suitable for distributed scalable setups. In particular, HDFS supports the following row-oriented formats (e.g., \textit{CSV} and \textit{Avro}) and columnar formats (e.g., \textit{ORC} and \textit{Parquet}).

\subsection{Bench-Ranking in a Nutshell}\label{sec:benchranking}

\begin{wrapfigure}[7]{r}{0.35\textwidth}
    \vspace{-40pt}
    \centering
	\includegraphics[width=0.6\linewidth]{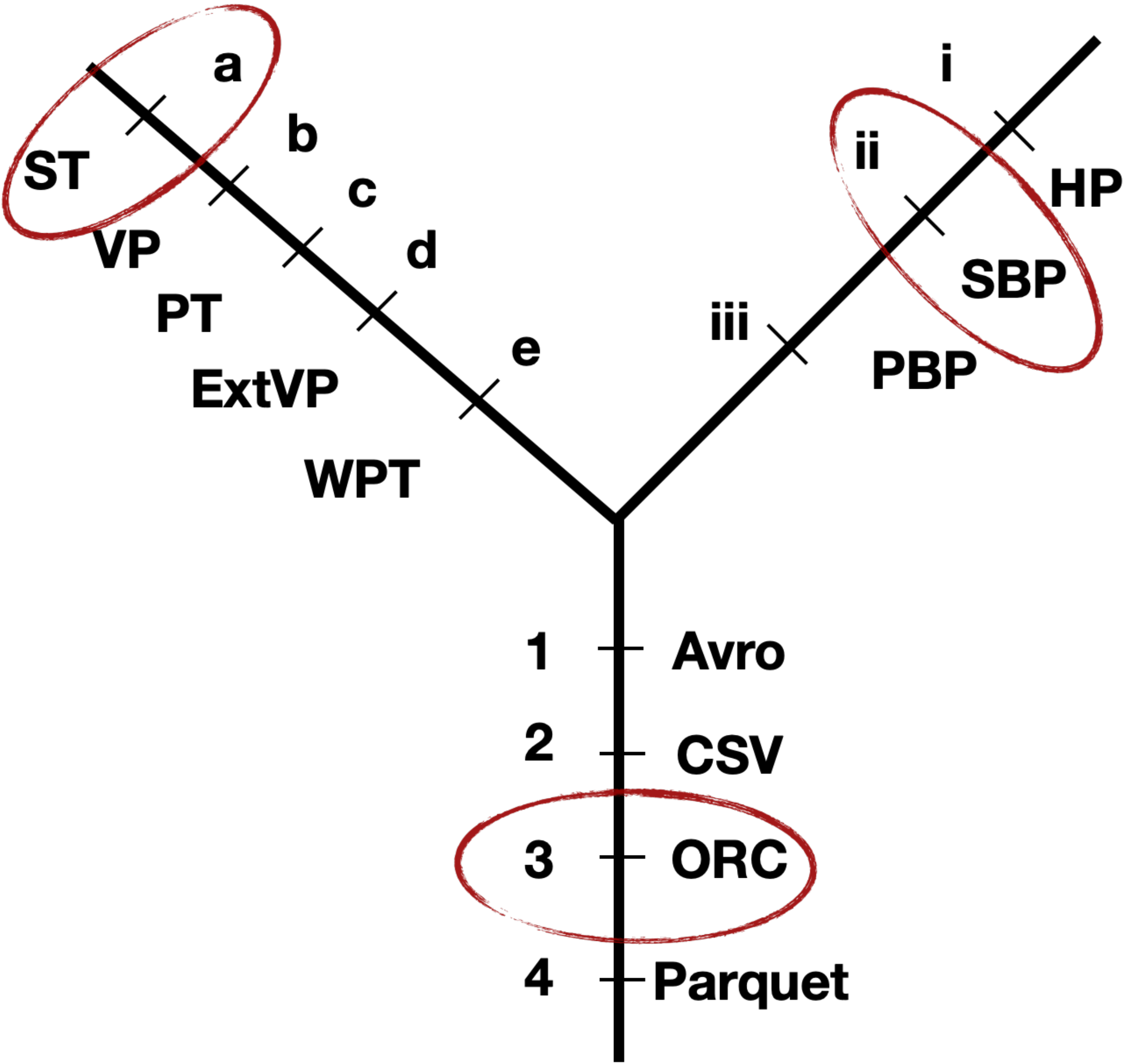}
	\caption{The configuration space $\mathcal{C}$.}
	\label{fig:experimentsarch}
\end{wrapfigure}

This section summarizes the concept of Bench-Ranking as a means for Prescriptive Performance Analysis. Bench-Ranking is based on \textit{three} fundamental notions, i.e., \textit{Configuration}, \textit{Ranking Function}, and \textit{Ranking Set}, defined below.

\begin{definition}\label{def:configcombin}
A configuration $c$ is a combination of experimental dimensions. The configuration space $\mathcal{C}$ is the Cartesian product of the possible configurations.
\end{definition}

In~\cite{ragabieee21}, we consider a three-dimensional configuration space, i.e., including \textit{relational schemas}, \textit{partitioning techniques} and \textit{storage formats}. Figure~\ref{fig:experimentsarch} shows the experimental space and highlights the example of the (\textit{a.ii.3}) configuration, which is akin to the Single Triples (\textit{ST}) schema, Subject-based Partitioning (\textit{SBP}) technique, and stored in the HDFS~\textit{(ORC)} storage file format. This naming convention guides configurations reading in the rest of the paper results (figures and tables).

\begin{definition}\label{def:rset}
A \textit{ranking set} $\mathcal{R}$ is an ordered set of elements ordered by a  rank score. The rank index $r_i$ is called the index of a ranked element $i$ within the ranking set $\mathcal{R}$, i.e., $\mathcal{R}$[$r_i$]=$i$. We denote with $\mathcal{R}^k$ the leftmost ($k$ top-ranked) subset of $\mathcal{R}$, and we denote with $\mathcal{R}_x$ the ranking set calculated according to the Rank score R$_x$. 
\end{definition}

\begin{definition}  \label{def:rfunction}
A ranking set is defined by a \textit{ranking function} $f_R: \mathcal{C} \rightarrow \mathcal{R}$ that associates a rank score to every element in $\mathcal{C}$.
\end{definition}

A valid example of a ranking score can be the time required for query executions by each of the selected configurations (see Table~\ref{tab:queryruntimes_rankings}). The \textit{ranking function} abstracts this notion (Definition~\ref{def:rfunction}). In~\cite{ragabieee21}, we consider a generalized version of the ranking function presented in~\cite{akhter2018empirical}, which calculates the rank scores for the configurations as follows:


\begin{table}[!t]
\scriptsize

\begin{minipage}{.5\linewidth}
  \centering
\begin{tabular}{|c|c|c|c|c||c|c|c|c|c|c|}
\hline
\backslashbox{Conf.}{Query} &
  \multicolumn{1}{l|}{Q1} &
  \multicolumn{1}{l|}{Q2} &
  \multicolumn{1}{l|}{...} &
  \multicolumn{1}{l||}{Q20} &
  \multicolumn{1}{l|}{Q1} &  
  \multicolumn{1}{l|}{Q2} &
  \multicolumn{1}{l|}{...} &  
  \multicolumn{1}{l|}{Q20}   \\ \hline
a.i.1   & 63.8   & 50.9   & ... &  13.5  & $41^{th}$ & $41^{th}$ & ...& $40^{th}$ \\ \hline
a.i.2   & 133.3   & 147.5  & ... & 46.8 & $44^{th}$ & $48^{th}$ & ...& $46^{th}$ \\ \hline
...     & ... & ... & ... & ... & ...  & ... & ... & ... \\ \hline
b.iii.4   & 24.8   & 25.2   & ...  & 7.5  & $11^{th}$ & $26^{th}$ & ...& $32^{th}$ \\ \hline
...     & ... & ... & ... & ... & ... & ... & ... & ... \\ \hline
\textbf{e.ii.4} & 205.2  & 162.3   & ...  & 26.6 & $60^{th}$ & $60^{th}$ & ...& $42^{th}$  \\ \hline
\end{tabular}
\caption{Configuration rankings by query execution time, e.g., (a.i.1) configuration is at $41^{th}$ rank in $Q1$.}
\label{tab:queryruntimes_rankings}
    \end{minipage}%
    \begin{minipage}{.5\linewidth}
      \centering
        
\begin{tabular}{c|cccccc}
\hline
\textbf{Schema} & \textbf{1$^{st}$} & \textbf{2$^{nd}$} & \textbf{3$^{rd}$} &  \textbf{4$^{th}$}  &  \textbf{5$^{th}$}  & \textbf{R} \\ \hline

\cellcolor[HTML]{D9D9D9}\textbf{ExtVP} & 6 & 6 & 8 &  0 & 0& \cellcolor[HTML]{8C872E}0.73 \\ \hline

\cellcolor[HTML]{D9D9D9}\textbf{PT} & 6 & 6 & 5 &  2 & 1& \cellcolor[HTML]{8C872E}0.68 \\ \hline

\cellcolor[HTML]{D9D9D9}\textbf{WPT} & 7 & 3 & 0 &  0 & 10& \cellcolor[HTML]{A36F16}0.46 \\ \hline

\cellcolor[HTML]{D9D9D9}\textbf{ST} & 1 & 3 & 4 &  9 & 3& \cellcolor[HTML]{A36F16}0.38 \\ \hline

\cellcolor[HTML]{D9D9D9}\textbf{VP} & 0 & 2 & 3 & 9 & 6 & \cellcolor[HTML]{A36F1F}0.26 \\ \hline

\end{tabular}
\caption{Example of \textit{Rank Scores}.}
\label{tab:formulaCalc}
\end{minipage} 
    
\end{table}
  
\begin{ceqn}
\label{eq:rankingscoreformula}\small
\begin{align}
  R=\sum_{r=1}^{d}\frac{O_{dim}(r)*(d-r)}{|Q|*(d-1)}, 0< R \leq 1
\end{align}
\end{ceqn}

In Equation (1), $R$ is the \textit{rank score} of the ranked dimension (i.e., relational schema, partitioning technique, storage format, or any other experimental dimensions). Such that, $d$ represents the total number of parameters (options) under that dimension (for instance \textit{five} in case of schemas, see Figure~\ref{fig:experimentsarch}), $O_{dim}(r)$ denotes the number of occurrences of the dimension being placed at the rank $r$ ($1^{st}$, ${2^{nd}}$,..
). Moreover, $|Q|$ represents the total number of queries. Rank scores define the \textit{Single-Dimensional} (SD) ranking criteria that help to provide a \textit{high-level} view of the system performance across a set of tasks (e.g., queries in a workload)~\cite{ragabieee21}. Table~\ref{tab:formulaCalc} shows a simple example of applying the above formula for computing the rank scores of the relational schema dimension. The ExtVP schema was placed in the \textit{"first"} rank \textit{(six)} times (i.e., performed the best in six queries), the \textit{"second"} \textit{(six)} times, the \textit{"third"} \textit{(eight)} times, the \textit{"fourth"} \textit{(zero)} times, and the last \textit{"fifth"} also \textit{(zero)} times. Thus, its overall ranking is $0.73$. In contrast, the VP performed the worst with a rank score of $0.26$. Intuitively, this means that the ExtVP schema in this example is the best (i.e., it has the highest rank score), and the VP schema is the worst-performing one.

Despite its generalization, Equation (1) is insufficient for ranking the configurations in a configuration set defined $\mathcal{C}$ when it counts multiple dimensions~\cite{ragabieee21}. The presence of \textit{trade-offs}~\cite{ragabSBD,ragabieee21} reduce the accuracy of these SD ranking functions. 
Thus, we introduced Multi-Dimensional ranking~\cite{ragabieee21} by approaching Bench-Ranking as a \textit{Multi-objective Optimization} problem. In particular, we adopt the \textit{Pareto frontier} optimization techniques~\footnote{Pareto frontier aims at finding a set of optimal solutions if no objective can be improved without sacrificing at least one other objective.}, implemented using \textit{Non-dominated Sorting Genetic Algorithm} (\textit{NSGA-II})~\cite{nsga2}, to optimize the experimental dimensions altogether and find the best-performing configuration in $\mathcal{C}$. 
%

Finally, our Bench-Ranking frameworks include \textit{two} evaluation metrics to assess the \textit{effectiveness} of the proposed ranking criteria. In particular, we consider a ranking criterion is \textit{good} if it does not suggest \textit{low-performing} configurations and if it minimizes the number of contradictions within an experimental setting. When it comes to PPA, practitioners are not interested in a configuration is being the fastest at answering any specific query in a workload as long as it is never the slowest at any of the queries. To this extent, we identified two evaluation metrics~\cite{ragabieee21}, i.e., 
\begin{inparaenum}[(i)]
    \item \textbf{Conformance}, which measures the \textit{adherence} of the \textit{top-ranked} configurations w.r.t actual query results (see Table~\ref{tab:queryruntimes_rankings}); 
    \item\textbf{Coherence} which measures the level of (dis)agreement between two ranking sets that use the \textit{same} ranking criterion across different experiments (e.g., different dataset sizes).
\end{inparaenum}

We calculate the conformance according to Equation (2) by positioning an element in a ranking set w.r.t the initial rank score. For instance, let's consider a ranking criterion $\mathcal{R}_x$ with the \textit{top-3} ranked configurations ($k=3$) are $\mathcal{R}_x^{k=3}$=\{d.ii.3, b.ii.2,\textbf{e.ii.4}\}, that overlap only with the \textit{bottom-3} ranked configurations ($h=3$) in one query $\mathcal{Q}_x$, as shown in Table~\ref{tab:queryruntimes_rankings}. That is, $\mathcal{Q}_x^{h=3}$=\{e.iii.1, e.iii.3, \textbf{e.iii.4}\}, i.e \textbf{$e.iii.4$} is in the $60^{th}$ position out of $60$ ranks/positions (i.e., the last rank). Thus, the Conformance of $(\mathcal{R}_x^3)=1-1/(20*3)$, when $k=3$, $h=3$, and $Q=20$.

\begin{ceqn}\label{eq:cirteriaaccuracyformula}
\begin{align}
A(\mathcal{R}^{k})=1-\sum _{i=0}^{|Q|}\sum _{j=0}^{k} \frac{\bar{A}(i,j)}{|Q|*k},~~~ \bar{A}(i,j)=\begin{cases}
 1 & \mathcal{R}^k[j] \in \mathcal{Q}i_h^{i}\\
 0 & otherwise
\end{cases} 
\end{align}
\end{ceqn}

For coherence, we employ \textit{Kendall's index}~\footnote{Kendall is a standard measure to compare the outcomes of ranking functions.} according to Equation (3), which counts the number of pairwise (dis)agreements between two ranking sets,  Kendall’s distance between two ranking sets $\mathcal{R}1$ and $\mathcal{R}2$, where $P$ is the set of \textit{unique} pairs of distinct elements. The larger the distance, the more \textit{dissimilar} the ranking sets are.  
\vspace{5pt}

\begin{minipage}{0.5\textwidth}
\[\bar{K}_{i, j}\left(\mathcal{R}1, \mathcal{R}2\right )= \begin{cases}
    0 &
    \mathcal{R}1[r^1_i]=
    \mathcal{R}2[r^2_i]= i  ~\wedge 
    \mathcal{R}1[r^1_j] \\& =
    \mathcal{R}2[r^2_j]= j ~\wedge \\
    & r^1_i-r^1_j = r^2_i - r^2_j\\
   1              & \text{otherwise}
\end{cases}\]
    \end{minipage}
\begin{minipage}{.5\textwidth}
\begin{ceqn}\label{eq:kendall}
\begin{align}
K\left(\mathcal{R}1, \mathcal{R}2\right)=\sum_{\{i, j\} \in P}\frac{\bar{K}_{i, j}\left(\mathcal{R}1, \mathcal{R}2\right)}{\mid P \mid}
\end{align}
\end{ceqn}
    \end{minipage}%
\vspace{3pt}

We assume that ranking sets have the same number of elements. For instance, the $K$ index between $\mathcal{R}1$=(a.ii.3,a.i.3,\textbf{b.ii.2}) and $\mathcal{R}2$=(a.ii.3,a.i.3,\textbf{c.iii.2}) for $100M$ and $500M$ is $0.33$, i.e., one disagreement out of three comparisons.

\vspace{-10pt}

\section{\pap}\label{sec:lib_req_arch}

In this section, we present the requirement analysis for \pap library, and describe its architecture. We elicit \pap's requirements based on the implementation challenges we discussed in the introduction and on the existing research efforts on benchmarking BD systems for processing and querying large RDF graphs~\cite{schatzle2016s2rdf,schatzle2014sempala,ragabieee21,ragabSBD,cossu2018prost}.

\subsection{Requirements} 


\begin{figure}[b]
\centering
\includegraphics[width=0.6\linewidth]{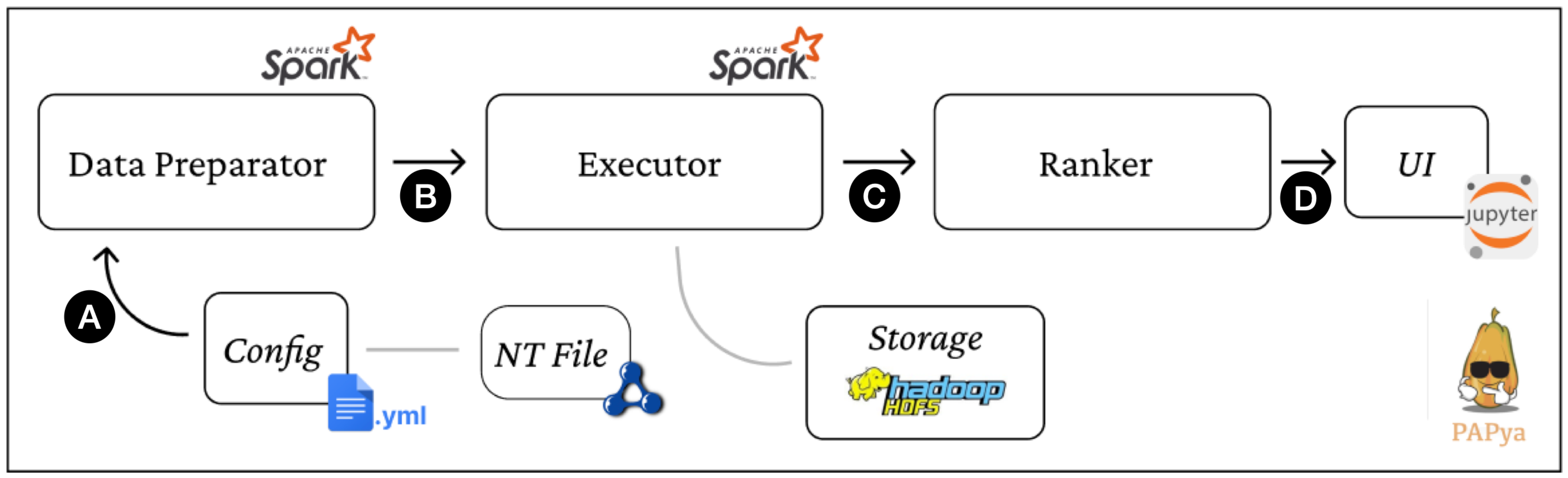}
\caption{Papaya architecture and workflow.}
\label{fig:papayarch}
\end{figure}

\begin{enumerate}[leftmargin=*, label=\textbf{R.\arabic*}]

    \item \label{req:br}\label{req:holistic}  \textbf{Support for PPA}: \pap shall support the necessary abstractions required to support PPA. Moreover, by default, it shall support existing Bench-Ranking techniques in~\cite{ragabieee21}.
    
    \item  \label{req:kpi} \textbf{Independence from the Key Performance Indicators (KPIs)}: \pap must enable PPA independently from the chosen KPI. In~\cite{ragabieee21}, we opted for query latency, yet one may need to analyze the performance in terms of other metrics (e.g., memory consumption).

    \item \label{req:dim} \textbf{Independence from Experimental Dimensions}: \pap must allow the definition of an arbitrary number of dimensions, i.e., allow the definition of \textit{n-dimensional} configuration space. 
    
    
    \item \label{req:viz} \textbf{Usability}: \pap supports decision-making, simplifying the performance data analytics. To this extent, data visualization techniques and a simplified API are both of paramount importance. 

    \item  \label{req:ext} \textbf{Flexibility and Extensibility}: \pap should be extensible both in terms of architecture and programming abstractions to adapt with adding/removing configurations, dimensions, workload queries, ranking methods, evaluation metrics, etc. It also should decouple data and processing abstractions to ease the integration of new components, tools, and techniques.
    
\end{enumerate}

\subsection{Architecture, Abstractions, and Internals}

\begin{table}[b]
\scriptsize
\begin{tabular}{l|c|l}
\hline
\rowcolor[HTML]{C0C0C0} 
\textbf{Challenges} & \textbf{Requirements} & \textbf{\pap Solutions} \\ \hline
\textbf{C.1}: Experiments Preparation & \textbf{R4, R5} & \begin{tabular}[c]{@{}l@{}}- Data Preparator that generates and loads graph data for distributed relational setups.\\ - User-defined \textit{YAML} configurations files.\end{tabular} \\ \hline
\textbf{C.2}: Portability and Usability & \textbf{R2, R3, R4, R5} & \begin{tabular}[c]{@{}l@{}}- Data Preperator prepares graphs data ready for processing with any arbitrary relational BD system.
\\ - Internals \& abstractions enable plugin-in new modules and programmable artifacts.\color{black}  
\\ - Checking performance replicability whilst configuration changes.  \end{tabular} \\ \hline
\textbf{C.3}: Flexibility and Extensibility & \textbf{R1, R3, R5} & \begin{tabular}[c]{@{}l@{}}- Allow adding/excluding experimental dimensions.\\ - Allow adding new ranking algorithms.\\ - Flexiblity in shortening and enlarging the configuration space.\end{tabular} \\ \hline
\textbf{C.4}: Compoundness and Complexity & \textbf{R1, R4} & \begin{tabular}[c]{@{}l@{}}- Interactive Jupyter Notebooks.\\ - Variety of data and ranking visualizations\end{tabular} \\ \hline

\end{tabular}
\caption{Summary of challenges and requirements mapping along with \pap solutions.}
\label{tab:challenge_req}
\vspace{-3pt}
\end{table}

This section presents the \pap's main components and shows how they fulfill the requirements. Table~\ref{tab:challenge_req} summarizes the requirements to challenges mappings alongside the \pap solutions.  
\pap allows its user to build an entire pipeline for querying big RDF datasets and analyzing the performance results. In particular, it facilitates building the experimental setting considering the configuration space (described in Definition \ref{def:configcombin}) specified by users. This entails preparing and loading the graph data in a \textit{user-defined relational} configuration space, then performing experiments (executing a query workload on top of a relational BD framework), and finally analyzing and providing prescriptions of the performance.

\begin{figure}[t!]
\centering
\includegraphics[clip,trim=0 0 0 0,width=0.75\linewidth]{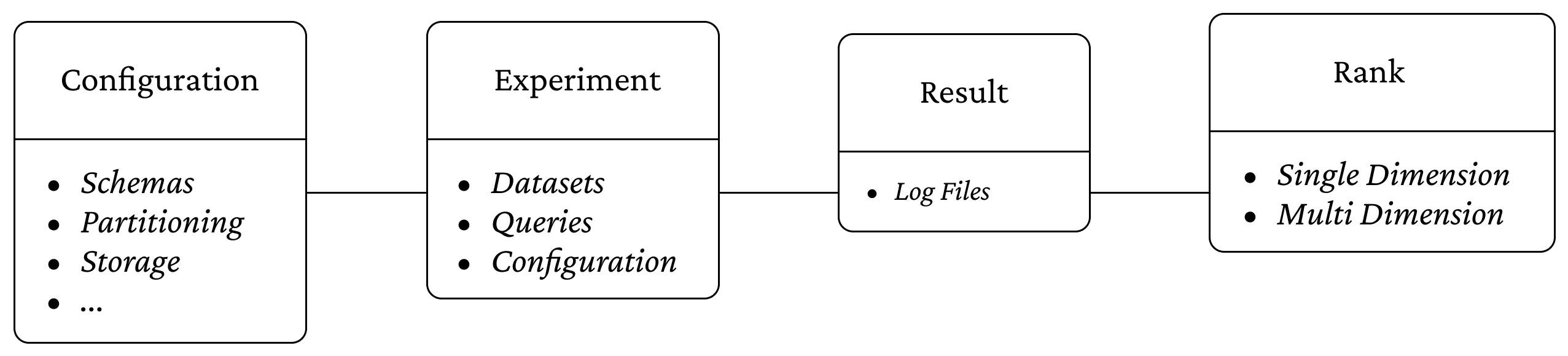}
\caption{\pap internal abstractions.}
\label{fig:abstractuml}
\vspace{-5pt}
\end{figure}

To achieve that, \pap includes \textit{three} core modules depicted in Figure~\ref{fig:papayarch}, i.e., the \textit{Data Preparator}, the \textit{Executor}, and the \textit{Ranker}. Moreover, \pap relies on few core abstractions depicted in Figure~\ref{fig:abstractuml}, i.e.,
\textit{Configuration}, \textit{Experiment}, \textit{Result}, and \textit{Rank}.
While detailing each module's functionalities, we introduce \pap workflow, which also appears in Figure~\ref{fig:papayarch}, starting with the input is a configuration file that points to the input \textit{N-Triple} file with the RDF graph (Figure~\ref{fig:papayarch} step (A)). 

The first actor in the pipeline is the \textit{Data Preparator} (DP), which prepares RDF graphs for relational processing. It takes as input a configuration file that includes experimental options of interest. The configuration file is represented by the \textit{Configuration} abstraction (see Figure~\ref{fig:abstractuml}), which enables extensibility (\ref{req:ext}). Specifically, the DP allows defining an arbitrary number of dimensions with as many options as specified (\ref{req:dim}). In particular, it considers the dimensions specified in~\cite{ragabieee21} (\ref{req:br}), i.e., relational schemas, storage format, and partitioning technique. Therefore, the DP automatically prepares (i.e., \textit{determines}, \textit{generates}, and \textit{loads}) the RDF graph dataset with the specified configurations that adapt with the relational processing paradigm to the storage destination (e.g., HDFS). More specifically, DP currently includes four relational schemas commonly used for RDF processing, i.e., (a) ST, (b) VP, (d) ExtVP, and (e) WPT~\footnote{Automating PT schema ("b" in Figure~\ref{fig:experimentsarch}) generation is not yet supported by the current \pap DP}. 
%
For partitioning, DP currently supports three partitioning techniques, i.e., (i) horizontal partitioning (HP), (ii) subject-based partitioning (SBP), and predicate-based partitioning (PBP). 
Last but not least, DP enables storing data using four HDFS file formats (i) \textit{CSV} or (ii) \textit{Avro}, which are row-oriented, and (iii) \textit{ORC} or (iv) \textit{Parquet}, which are column-oriented storage formats~\cite{ivanov2019impact}.
The DP interface is generic, and the generated datasets are~\textit{agnostic} to the underlying relational system. The DP prepares RDF \textit{graph} data for processing with different \textit{relational} BD systems, especially SQL-on-Hadoop systems, e.g., SparkSQL, Hive, Impala. Seeking scalability, the current DP implementation relies on \textit{SparkSQL}, which allows implementation of RDF relational schema generation using the SQL transformations. 
%
Notably, \textit{Apache Hive} or \textit{Apache Impala} could be potential candidates for an alternative implementation executors. However, SparkSQL also supports different partitioning techniques and multiple storage formats, making it ideal for our experiments~\cite{ragabieee21}. 

\begin{figure}[t!]
    \centering
	\includegraphics[width=0.75\linewidth]{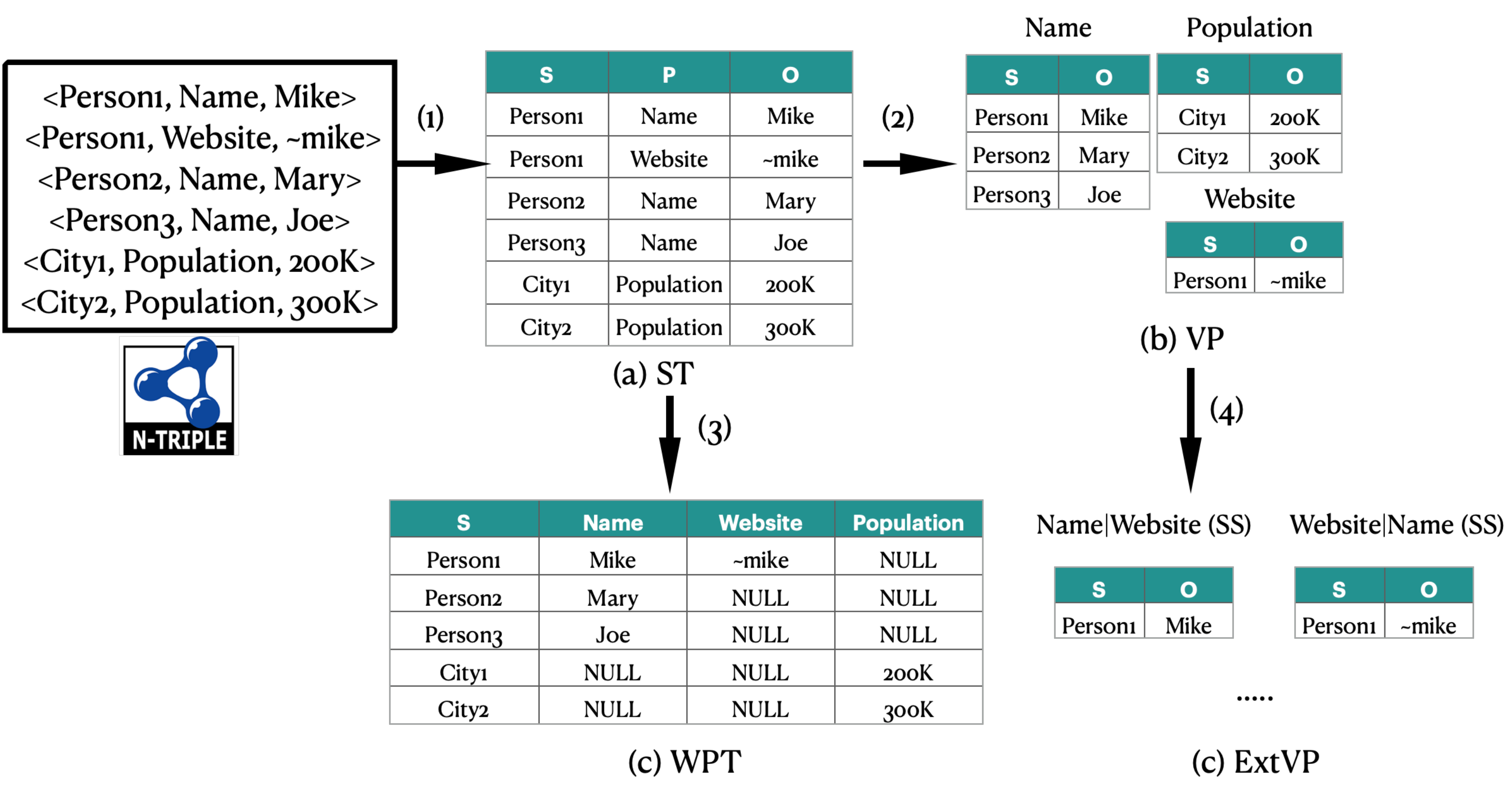}
	\caption{RDF relational schema transformations in \pap Data Preparator.}
	\label{fig:schemagen}
	\vspace{-10pt}
\end{figure}

Figure~\ref{fig:schemagen} shows sample of schema generation in \pap DP component. First, the DP transforms the input RDF graph (\textit{N-Triples} file(s)) into an ST table schema (i.e., Figure~\ref{fig:schemagen} Step(1)), and then other schemas are generated using parameterized SQL queries~\footnote{schema SQL-based transformations are kept in the DP module on \pap's GitHub repository due to space limits.}. For instance, the VP and WPT schemas are generated using SQL queries given the ST table as a parameter (i.e., Figure ~\ref{fig:schemagen} Step(2), and (3), respectively). While, the ExtVP schema generation relies on VP tables to exist first (i.e., Step(4) in Figure ~\ref{fig:schemagen}). 

The \textbf{Executor} is the following module in \pap workflow, which is the system that is subject to experimentation (see Figure~\ref{fig:papayarch} step B). For instance, in~\cite{ragab2021depth,ragabieee21,schatzle2016s2rdf,cossu2018prost} the considered system is Apache SparkSQL. The executor offers an abstract API to be extended (\ref{req:ext}).
In practice, it (i) starts the execution pipeline in the external system, (ii) it collects the performance logs (\ref{req:kpi}), and currently, it persists them on a file system, e.g., HDFS. The Executor expects a set of experiments to run defined in terms of (i) a set of queries,  (ii) an RDF dataset (size), and (iii) a configuration (defined in Definition~\ref{def:configcombin}). The \textit{Experiment} abstraction is defined in Figure~\ref{fig:abstractuml}. We decide to wrap the running experiments in a \textit{SparkSQL-based} executor (\textit{SparkExecutor}). The experiment specifications (alongside the configurations) are passed to this wrapper as parameters. It is worth mentioning that the Executor assumes the query workload is available in the form of the \textit{SQL} queries. In the current stage, \pap does not support SPARQL query translation nor SQL query mappings.

The results logs are then loaded by the \textbf{Ranker} component into \textit{python Dataframes} to make them available for analysis (see Figure~\ref{fig:papayarch} step (C)). The Ranker reduces the time required to calculate the rankings, obtain useful data visualizations, and determine the \textit{best-performing} configurations while checking the performance replicability. To fulfill \ref{req:kpi}, i.e., the Ranker component operates over a log-based structure whose schema shall be specified by the user in the input configurations. Moreover, to simplify the usage and the extensibility (\ref{req:ext}), we decoupled the performance analytics (e.g., ranking calculation) from its definitions and visualization. 
In particular, the \textit{Rank} class abstraction reflects on the \textit{ranking function} (Definition~\ref{def:rfunction}) that takes as input data elements and returns a \textit{ranking set}. To fulfill \ref{req:viz}, a default data visualization for the rank shall be specified. However, this is left for the user to specify due to the specificity of the visualization.

The Rank call allows defining additional ranking criteria (\ref{req:ext}). In addition, to fulfill \ref{req:br}, \pap already implements SD as well as \textit{Multi-Dimensional} (MD) criteria specified in~\cite{ragabieee21}.  


\pap allows its users to interact with the experimental environment (\ref{req:ext}) using \textit{Jupyter Notebook}. To facilitate the analysis, it integrates data visualization (\ref{req:viz}) that can aid decision-making. Thus, the~\textit{Rank} class includes a specific method to implement, where to specify the default visualization.

Finally, to evaluate the raking criteria, we introduced in Section~\ref{sec:benchranking} the notions of coherence and conformance (\ref{req:br}). Ranking criteria evaluation metrics are employed to select which ranking criterion is \textit{“effective”} (i.e., if it is not suggesting low-performing configurations). In our experiments, we use such metrics by looking at all ranking criteria and comparing them with the results across different scales, e.g., dataset sizes ($100M$, $250M$, and $500M$). 
Notably, to minimize the dependencies, we implemented the ranking algorithms and evaluation metrics from scratch.

\begin{table}[t!]
\centering
\scriptsize
\label{tab:SDbestconfs}
\begin{tabular}{|c|c|c|c|l|l|l|c|c|c|c|l|l|l|} 
\hline

\multirow{2}{*}{$D_i$}     & \multicolumn{6}{c|}{{\cellcolor[rgb]{0.851,0.851,0.851}}WatDiv$_{mini}$}                                                                          &  & \multicolumn{6}{c|}{{\cellcolor[rgb]{0.851,0.851,0.851}}WatDiv$_{full}$}                                                                          \\ 
\cline{2-7}\cline{9-14}
                        & \textbf{$\mathcal{R}_f^3$} & \textbf{$\mathcal{R}_p^3$} & \textbf{$\mathcal{R}_s^3$} & Pareto$_{Q}$ & Pareto$_{Agg}$ & \textbf{\textbf{$\mathcal{R}_{ta}^3$}} &  & \textbf{$\mathcal{R}_f^3$} & \textbf{$\mathcal{R}_p^3$} & \textbf{$\mathcal{R}_s^3$} & Pareto$_{Q}$ & Pareto$_{Agg}$ & \textbf{\textbf{$\mathcal{R}_{ta}^3$}}  \\ 
\cline{1-7}\cline{9-14}
\multirow{3}{*}{$100M$} & a.ii.3  & a.ii.3  & c.i.4 &  c.ii.2       & a.ii.3         & a.ii.3  & & a.ii.3 & c.ii.2  & d.iii.4 & d.iii.2        & c.ii.2         & c.ii.2                                                              \\ 
\cline{5-7}\cline{12-14}
                        & b.ii.2   & a.ii.4                    & c.ii.2 & c.i.2       & c.ii.2         & c.ii.2 & & a.i.3  & b.iii.3  & d.iii.1 & d.iii.3       & d.iii.3         & d.iii.3                                                                        \\ 
\cline{5-7}\cline{12-14}
                        & a.i.3  & a.ii.2  & c.i.3   & b.ii.2   & b.ii.2         & b.ii.2  & & b.ii.2  & c.ii.1  & d.iii.3 & d.iii.4       & b.ii.2         & b.ii.2                                                                         \\ 
\cline{1-7}\cline{9-14}
\multirow{3}{*}{$250M$} &  a.ii.3 & a.ii.3  & c.i.4 & c.i.4       & c.i.4         & c.i.4  & & a.ii.3 & a.ii.1  & d.iii.4 & d.iii.2       & d.iii.3         & d.iii.3                                                                       \\ 
\cline{5-7}\cline{12-14}
                        & a.i.3   & a.ii.4  & c.i.3 & c.ii.2        & b.ii.2          & b.ii.2  & & a.i.3  & d.iii.3  & d.iii.3 & d.iii.3       & c.i.4         & c.i.4                                                                          \\ 
\cline{5-7}\cline{12-14}
                        & c.i.4                    & a.ii.2                      & c.i.2                      & c.i.2       & a.ii.3         & a.ii.3 & & b.iii.2                     & a.ii.2  & d.iii.2  & c.i.4       & a.iii.4         & a.iii.4                                                                        \\ 
\cline{1-7}\cline{9-14}
\multirow{3}{*}{$500M$} & a.ii.3  & a.ii.3                    & c.ii.4 & c.ii.3       & b.ii.2         & b.ii.2 & & a.ii.3  & c.ii.2  & d.iii.2  & d.ii.2        & d.ii.2         & c.ii.3                                                            \\ 
\cline{5-7}\cline{12-14}
                         & a.i.3   & a.ii.4                    & c.i.4 & c.ii.4       & c.ii.3         & c.ii.3 & & a.i.3 & a.iii.2  & d.iii.4 & c.ii.3       & c.ii.3         & d.ii.2                                                                        \\ 
\cline{5-7}\cline{12-14}
                         & c.i.3                    & a.ii.2  & c.i.3                     & c.i.3      & c.ii.4        & c.ii.4  & & c.iii.2                     & a.iii.4  & d.iii.1 & a.iii.4       & a.iii.4         & a.iii.4                                                                      \\
\hline
\end{tabular}

\caption{WatDiv best-performing (\textit{Top-3}) configurations according to the SD and MD ranking criteria.}


\label{tab:results_watdiv}
\end{table}

\vspace{-10pt}
\section{\pap in Practice}\label{sec:bench_in_use}

\begin{wrapfigure}[20]{r}{.57\textwidth}
\vspace{-20pt}
\begin{minipage}{0.57\textwidth}
\begin{lstlisting}[language=python,frame=single, caption={Experiment design example in \pap.},label={lst:expdesign},captionpos=b, basicstyle=\scriptsize]
from papya import Configuration, SparkExecutor
from papya import data_preparator as dp
#Configurations
confs = Configuration({ (*\label{ln:conf1}*)
"schemas":["ST","VP","WPT","ExtVP"], 
"partition":["HP,SBP","PBP"], 
"storage":["CSV","Avro","Parquet", "ORC"]}) 
#Executor (*\label{ln:conf2}*)
Q=[q1,q2,..,q20]
exp = dp.experiment(dataset="100M", Q, confs) (*\label{ln:experiment}*)
spe = SparkExecutor(master="local[*]") (*\label{ln:executor}*)
res=spe.run(exp,runs=5,dataPath="hdfs:..",logsPath="hdfs:..")
#Bench-Ranking  (*\label{ln:br}*)
#(1) SD-Ranking Criteria
schemaSDRanks = SDRank(res,dim=list(conf.keys())[0], (*\label{ln:sdr}*)
q=len(Q), d=len(list(conf.values())[0]) #schema SDranking 
partitioningSDRanks = ... #partitioning SDRanking
storageSDRanks = ... #storage SDRanking
#(2) MD-Ranking (Pareto)
paretoFronts_Q=MDRankPareto(ds="100M",dims=Q) (*\label{ln:mdrq}*)
paretoFronts_Agg=MDRankPareto(ds="100M"       (*\label{ln:mdragg}*)
    ,dims=[schemaSDRanks,partitioningSDRanks,storageSDRanks])
#Visualization  (*\label{ln:viz}*)
SDRank.plot(schemaSDRanks) #plot SDRanking for schema
MDRrankPareto.plot(paretoFronts) #plot MD-Ranking Pareto
#Ranking Criteria evaluation (*\label{ln:valid}*)
conf=Ranker.conformance(schemaSDRanks,q=20,k=3, h=45) 
coh=Ranker.coherence(schemaSDRanks_100M, schemaSDRanks_250M) 
\end{lstlisting}
\end{minipage}
\end{wrapfigure}

In this section, we explain how to use \pap in practice, showcasing its functionalities with a focus on performance data analysis, flexibility, and visualizations. 
In particular, we design our experiments in terms of (i) a set of SPARQL queries that we manually translated into SQL accordingly with the different relational schemas
, (ii) RDF datasets of different sizes automatically prepared using our~\textit{Spark-based DataPreparator}, and (iii) a configuration based on three dimensions as in~\cite{ragabieee21}, i.e., schema, partitioning techniques, and storage formats. 

In Bench-Ranking experiments~\cite{ragabieee21}, we used the \textit{SP$^2$B}~\cite{schmidt2009sp} benchmark datasets. In this paper, we aim to use a different benchmark to check the robustness of \pap Bench-Ranking criteria and their evaluation metrics. Thus, we present the results of experiments opting for the \textit{WatDiv} benchmark~\cite{alucc2014diversified}~\footnote{Nonetheless, seeking conciseness, we keep \textit{SP$^2$B} results on the GitHub repository.}. \textit{WatDiv} benchmark includes a data generator and a query workload with various graph patterns, SPARQL query shapes, and selectivities that make the analysis \textit{non-trivial}. 
In our experiments, which are based on the average results of \textit{five} runs~\footnote{Benchmarks' query workload (in SQL) and experiments runtimes: \url{https://datasystemsgrouput.github.io/SPARKSQLRDFBenchmarking}}, we measure the performance of SparkSQL as a BD relational engine in terms of query \textit{latency}. However, alternative KPIs, e.g., memory consumption, could be easily used in \pap.

\begin{table}[t!]
\scriptsize
\begin{tabular}{|l|cccccc|c|cccccc|}
\hline
 &
  \multicolumn{6}{c|}{\cellcolor[HTML]{9B9B9B}WatDivMini} &
  \cellcolor[HTML]{FFFFFF} &
  \multicolumn{6}{c|}{\cellcolor[HTML]{9B9B9B}WatDifull} \\ \cline{2-7} \cline{9-14} 
 &
  \multicolumn{3}{c|}{\cellcolor[HTML]{C0C0C0}Conformance} &
  \multicolumn{3}{c|}{\cellcolor[HTML]{C0C0C0}Coherence} &
   &
  \multicolumn{3}{c|}{\cellcolor[HTML]{C0C0C0}Conformance} &
  \multicolumn{3}{c|}{\cellcolor[HTML]{C0C0C0}Coherence} \\ \cline{2-7} \cline{9-14} 
\multirow{-3}{*}{} &
  \multicolumn{1}{l|}{D1} &
  \multicolumn{1}{l|}{D2} &
  \multicolumn{1}{l|}{D3} &
  \multicolumn{1}{l|}{D1-D2} &
  \multicolumn{1}{l|}{D1-D3} &
  \multicolumn{1}{l|}{D2-D3} &
  \multicolumn{1}{l|}{} &
  \multicolumn{1}{l|}{D1} &
  \multicolumn{1}{l|}{D2} &
  \multicolumn{1}{l|}{D3} &
  \multicolumn{1}{l|}{D1-D2} &
  \multicolumn{1}{l|}{D1-D3} &
  \multicolumn{1}{l|}{D2-D3} \\ \cline{1-7} \cline{9-14} 
$\mathcal{R}_s$ &
  \multicolumn{1}{c|}{88.33\%} &
  \multicolumn{1}{c|}{91.67\%} &
  \multicolumn{1}{c|}{93.33\%} &
  \multicolumn{1}{c|}{0.09} &
  \multicolumn{1}{c|}{0.12} &
  0.06 &
   &
  \multicolumn{1}{c|}{96.00\%} &
  \multicolumn{1}{c|}{94.00\%} &
  \multicolumn{1}{c|}{93.00\%} &
  \multicolumn{1}{c|}{0.1} &
  \multicolumn{1}{c|}{0.13} &
  0.07 \\ \cline{1-7} \cline{9-14} 
$\mathcal{R}_p$ &
  \multicolumn{1}{c|}{38.33\%} &
  \multicolumn{1}{c|}{13.33\%} &
  \multicolumn{1}{c|}{5.00\%} &
  \multicolumn{1}{c|}{0.14} &
  \multicolumn{1}{c|}{0.14} &
  0.14 &
   &
  \multicolumn{1}{c|}{73.00\%} &
  \multicolumn{1}{c|}{39.00\%} &
  \multicolumn{1}{c|}{36.00\%} &
  \multicolumn{1}{c|}{0.09} &
  \multicolumn{1}{c|}{0.28} &
  0.17 \\ \cline{1-7} \cline{9-14} 
$\mathcal{R}_f$ &
  \multicolumn{1}{c|}{63.33\%} &
  \multicolumn{1}{c|}{46.67\%} &
  \multicolumn{1}{c|}{35.00\%} &
  \multicolumn{1}{c|}{0.16} &
  \multicolumn{1}{c|}{0.39} &
  0.3 &
   &
  \multicolumn{1}{c|}{64.00\%} &
  \multicolumn{1}{c|}{32.00\%} &
  \multicolumn{1}{c|}{43.00\%} &
  \multicolumn{1}{c|}{0.14} &
  \multicolumn{1}{c|}{0.25} &
  0.15 \\ \cline{1-7} \cline{9-14} 
Pareto$_{Q}$ &
  \multicolumn{1}{c|}{95.00\%} &
  \multicolumn{1}{c|}{98.33\%} &
  \multicolumn{1}{c|}{95.00\%} &
  \multicolumn{1}{c|}{0.14} &
  \multicolumn{1}{c|}{0.25} &
  0.14 &
   &
  \multicolumn{1}{c|}{92.00\%} &
  \multicolumn{1}{c|}{98.00\%} &
  \multicolumn{1}{c|}{98.00\%} &
  \multicolumn{1}{c|}{0.1} &
  \multicolumn{1}{c|}{0.15} &
  0.07 \\ \cline{1-7} \cline{9-14} 
Pareto$_{Agg}$ &
  \multicolumn{1}{c|}{88.33\%} &
  \multicolumn{1}{c|}{73.33\%} &
  \multicolumn{1}{c|}{93.33\%} &
  \multicolumn{1}{c|}{0.16} &
  \multicolumn{1}{c|}{0.25} &
  0.2 &
   &
  \multicolumn{1}{c|}{87.00\%} &
  \multicolumn{1}{c|}{71.00\%} &
  \multicolumn{1}{c|}{76.00\%} &
  \multicolumn{1}{c|}{0.17} &
  \multicolumn{1}{c|}{0.24} &
  0.15 \\ \cline{1-7} \cline{9-14} 
$\mathcal{R}_{ta}$ &
  \multicolumn{1}{c|}{88.33\%} &
  \multicolumn{1}{c|}{73.33\%} &
  \multicolumn{1}{c|}{93.33\%} &
  \multicolumn{1}{c|}{0.2} &
  \multicolumn{1}{c|}{0.24} &
  0.17 &
   &
  \multicolumn{1}{c|}{89.00\%} &
  \multicolumn{1}{c|}{84.00\%} &
  \multicolumn{1}{c|}{76.00\%} &
  \multicolumn{1}{c|}{0.15} &
  \multicolumn{1}{c|}{0.22} &
  0.13 \\ \hline
\end{tabular}

\caption{Ranking \textit{Coherence} (Kendall distance, the lower the better) \& \textit{Conformance} across \textit{WatDiv} datasets (D1=100M, D2=250M, D3=500M).}
\label{tab:eval_metrics_watDiv_mini_full}

\end{table}

Listing~\ref{lst:expdesign} shows a full example of \pap pipeline, starting by deciding the configurations (in terms of three dimensions and their options, e.g., list of schemas, partitioning techniques, storage formats to prepare, load, and benchmark) (Listing~\ref{lst:expdesign} lines~\ref{ln:conf1}-\ref{ln:conf2}). Then, an experiment is set up for running, defining the dataset size (e.g., "100M" triples), a list of queries to execute or exclude from the workload, and the configurations (Listing~\ref{lst:expdesign} line~\ref{ln:experiment}). An executor is defined for running the experiment along with the number of times experiments will be run (Listing~\ref{lst:expdesign} line~\ref{ln:executor}). The results (runtime logs) are kept in log files in a specified path (e.g., HDFS or a local disk). 
The \textit{Bench-Ranking} phase starts when we have the results in logs (Listing~\ref{lst:expdesign} line~\ref{ln:br})~\footnote{It is worth noting that the performance analyses, e.g., Bench-Ranking, could start directly if the performance data (logs) are already present.}. For instance, we call the \textit{SDRank} (Listing~\ref{lst:expdesign} line~\ref{ln:sdr}) for calculating rank scores for the "schema" dimension, alongside specifying the number of queries ("$q$"), and number of options under this dimension ("d" in Equation (1) ).  
The MD-Ranking(i.e., Pareto fronts) is applier in two ways. The first one is called Pareto$_Q$ (Listing~\ref{lst:expdesign} line~\ref{ln:mdrq}), which applies the \textit{NSGA-II} algorithm by considering the ranking sets obtained while sorting each query results individually. Using the first method, the algorithm aims at \textit{minimizing} the query runtimes across all dimensions. The second one is called the Pareto$_{Agg}$ (Listing~\ref{lst:expdesign} line~\ref{ln:mdragg}), which operates on the SD ranking criteria. By using the second method, the algorithm aims to \textit{maximize} the rank scores of the three SD-ranking criteria altogether, i.e., $R_s$, $R_p$, and $R_f$. 
%
The user can plot the SD rank scores and the MD Pareto ranking criterion (Listing~\ref{lst:expdesign} line~\ref{ln:viz}). In addition, the user can evaluate the effectiveness of the ranking criterion using conformance and coherence metrics (Listing~\ref{lst:expdesign} line~\ref{ln:valid}).

Table~\ref{tab:results_watdiv} shows the \textit{top-3} ranked configuration according to the various ranking criteria, i.e., Single-Dimension and Multi-Dimensional (Pareto) for the WatDiv datasets (i.e., 100M, 250M, 500M triples). In addition, Table~\ref{tab:eval_metrics_watDiv_mini_full} provides the ranking evaluation metrics (calculated according to Equations (2) and (3) 
). 


\subsection{Rich Visualizations}

\begin{figure}[p!]
\centering 
\begin{subfigure}{0.33\textwidth}
  \includegraphics[trim=0 0 0 0,clip,width=0.75\linewidth]{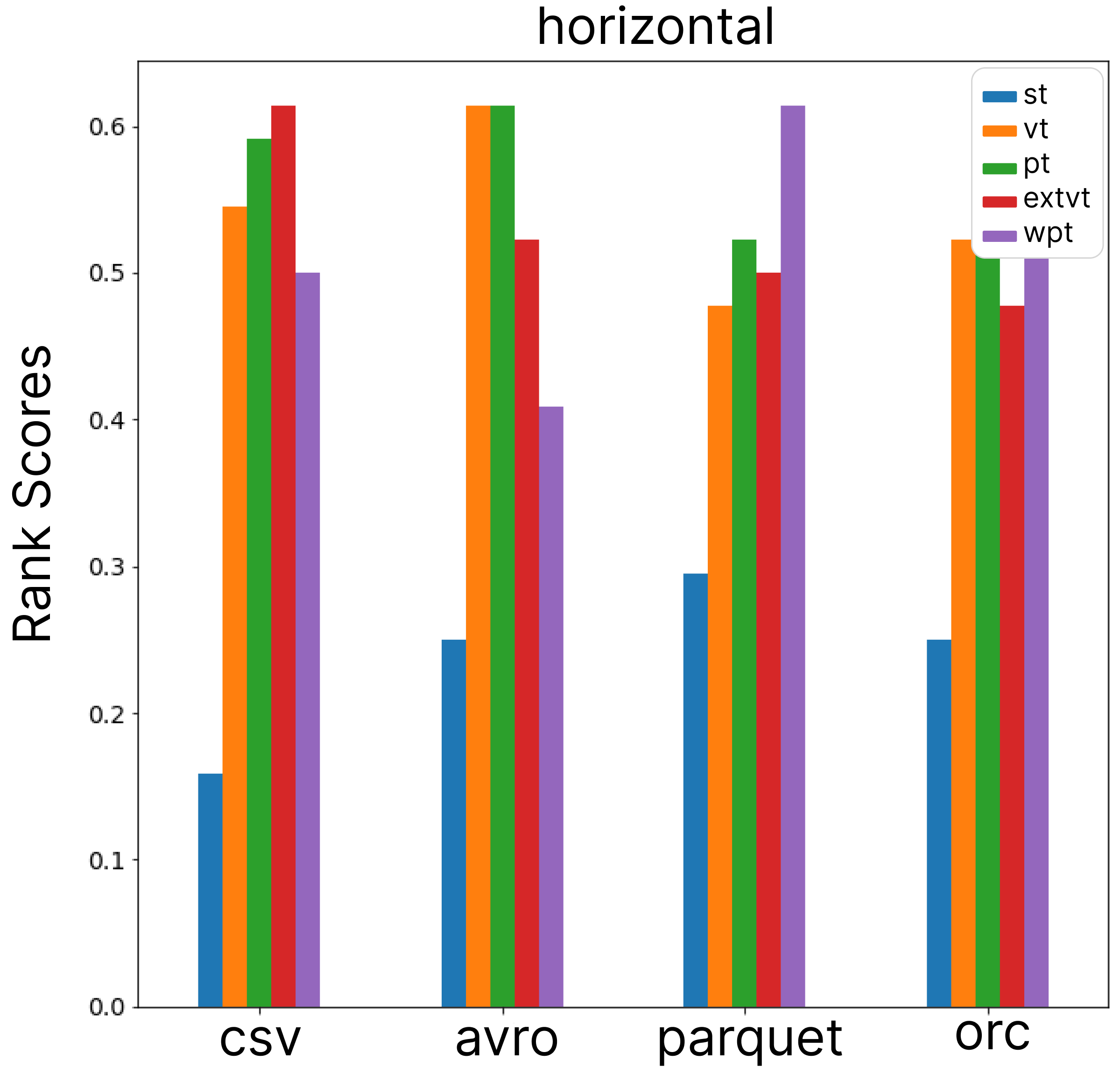}\quad
  \caption{SD Schema Ranks}
  \label{fig:1_sd_schema}
\end{subfigure}\hfil 
\medskip
\begin{subfigure}{0.33\textwidth}
  \includegraphics[trim=0 0 0 0,clip,width=0.75\linewidth]{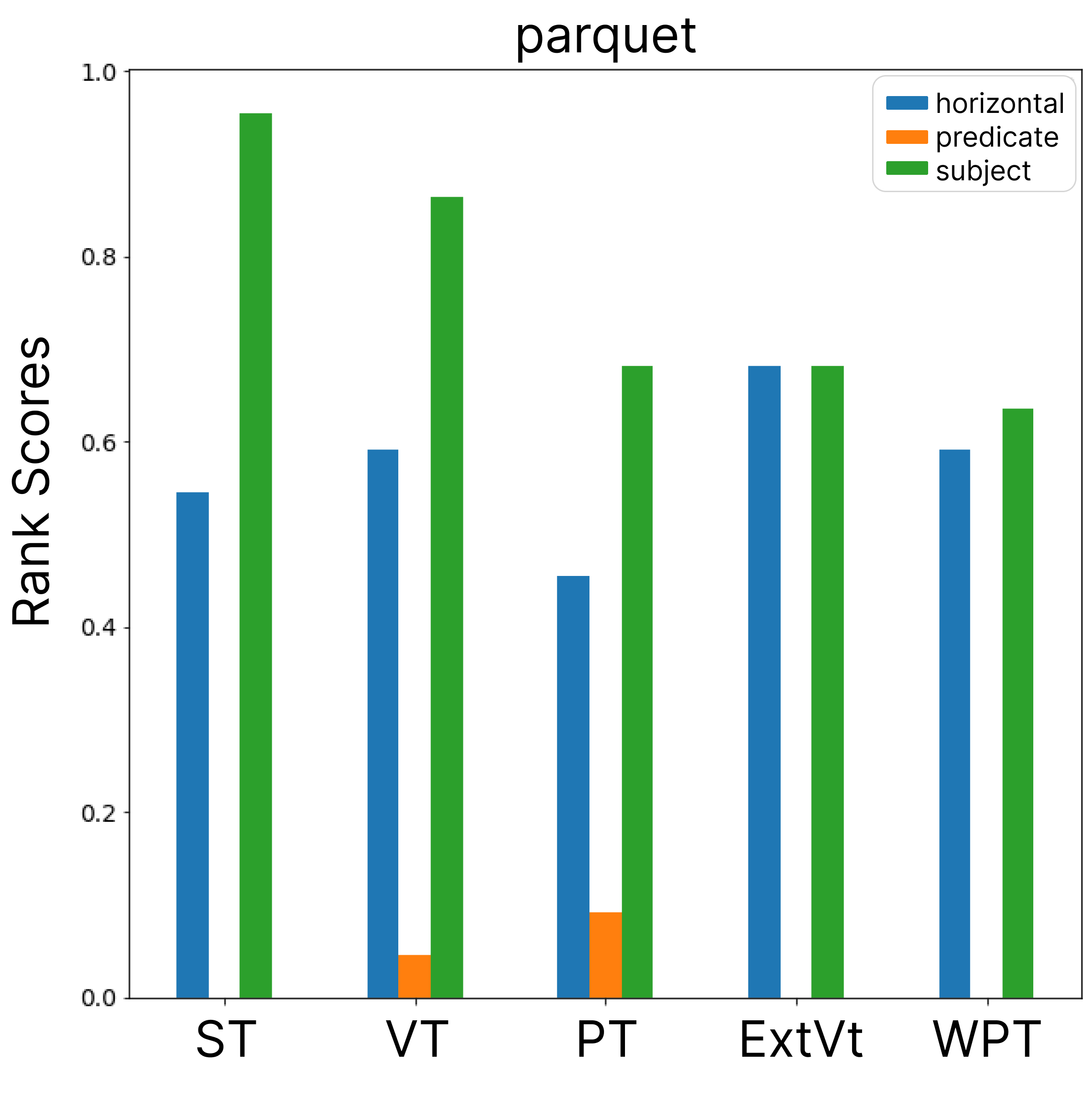}\quad
  \caption{SD Partition Ranks}
  \label{fig:2_sd_aprt}
\end{subfigure}\hfil 
\begin{subfigure}{0.33\textwidth}

\includegraphics[trim=0 0 0 0,clip,width=0.75\linewidth]{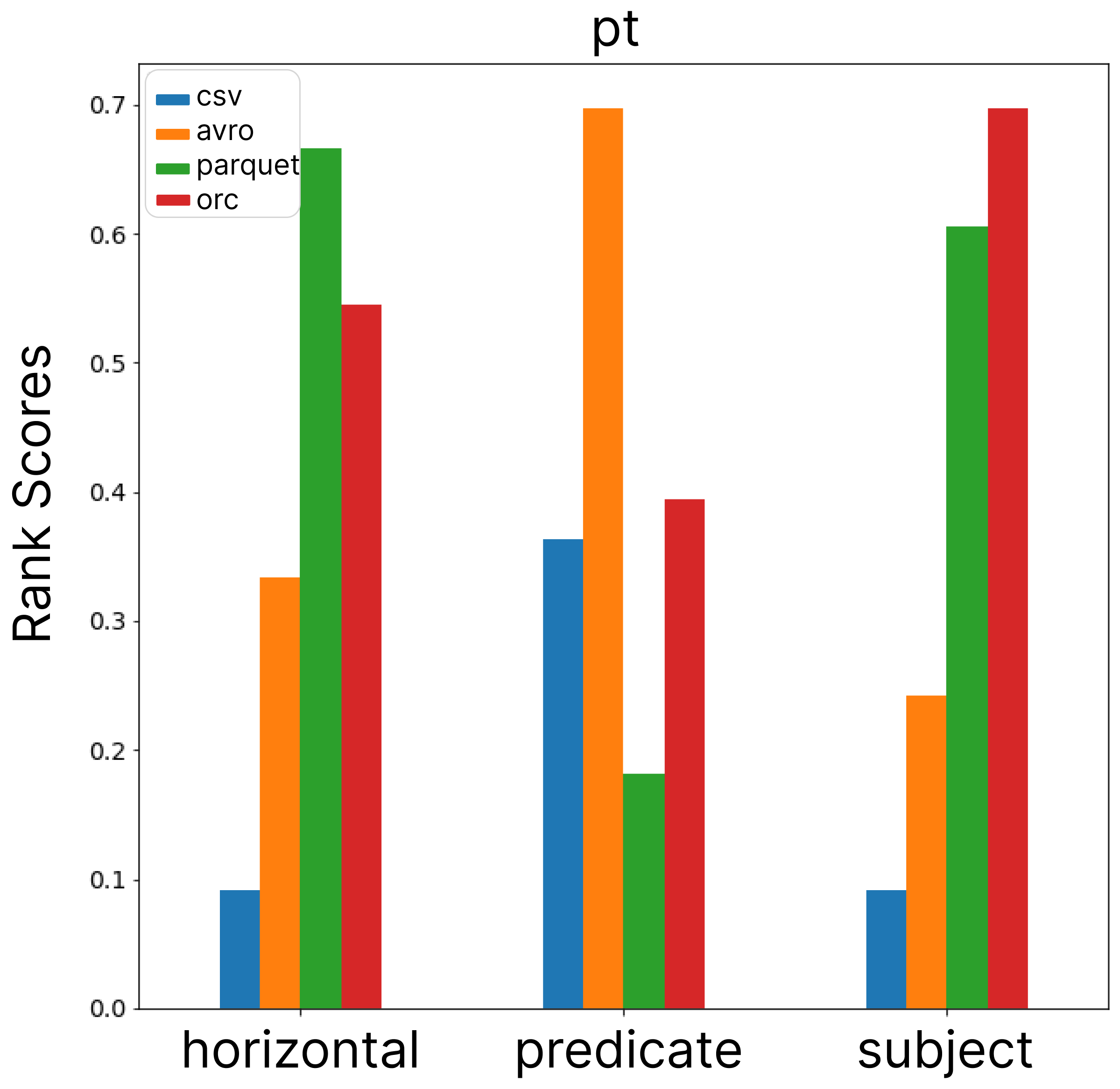}\quad
 \caption{SD Storage Ranks}
  \label{fig:3_sd_storage}
\end{subfigure}\hfil 
\vspace{-10pt}
\caption{Examples on SD Rank Scores over different dimensions (100M), the higher the better.}
\label{fig:sd}

    \centering 
\begin{subfigure}{0.32\textwidth}
  \includegraphics[trim=0 0 0 0,clip,width=0.75\linewidth]{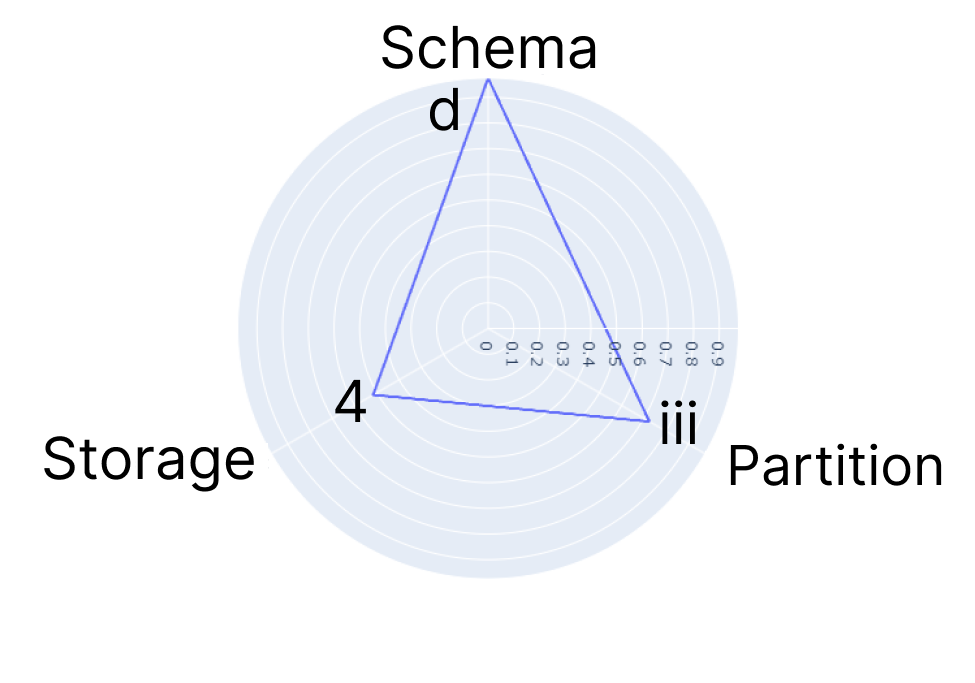}\quad
  \vspace{-5pt}
  \caption{Ranking by Schema}
  \label{fig:1_triangle}
\end{subfigure}\hfil 
\medskip
\begin{subfigure}{0.32\textwidth}
  \includegraphics[trim=0 0 0 0,clip,width=0.75\linewidth]{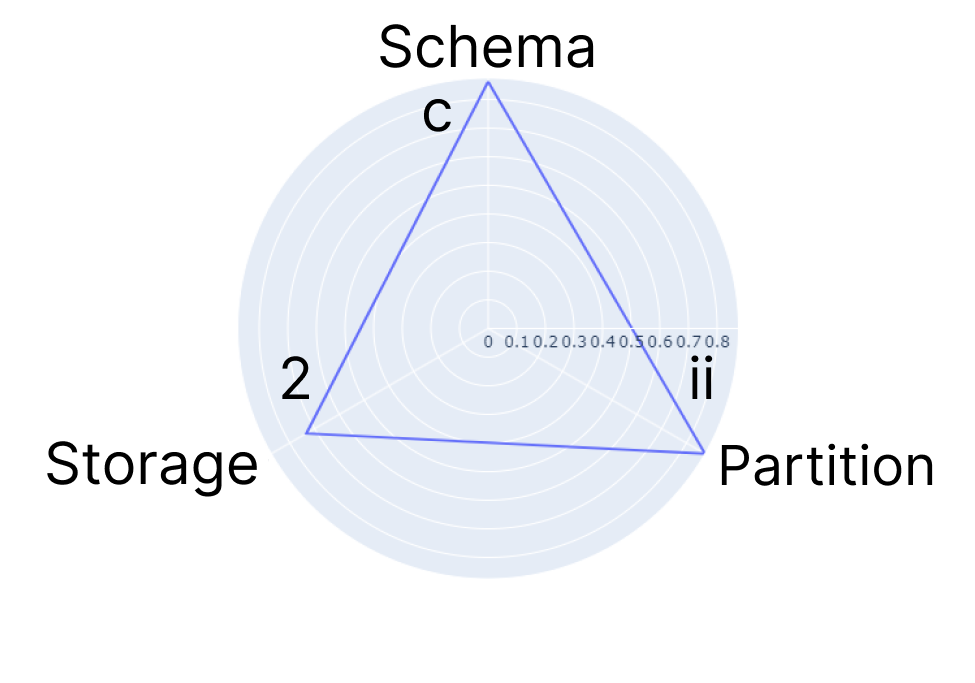}\quad
  \vspace{-5pt}
  \caption{Ranking by Partition}
  \label{fig:3_triangle}
\end{subfigure}\hfil 
\begin{subfigure}{0.32\textwidth}

\includegraphics[trim=0 0 0 0,clip,width=0.75\linewidth]{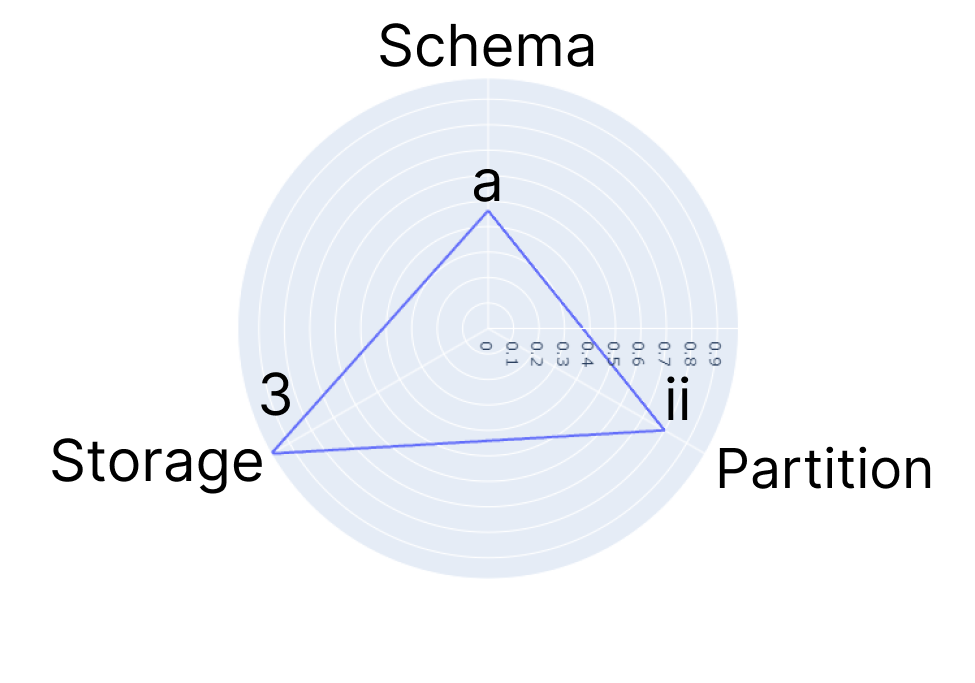}\quad
   \vspace{-5pt}
   \caption{Ranking by Storage}
  \label{fig:2_triangle}
\end{subfigure}\hfil 
\vspace{-15pt}
\caption{Dimensions trade-offs using single-dimensional ranking ($R_s$,$R_f$, and $R_p$).}
\label{fig:triangles}

    \centering 
\begin{subfigure}{0.5\textwidth}
  \includegraphics[trim=0 0 0 0,clip,width=0.7\linewidth]{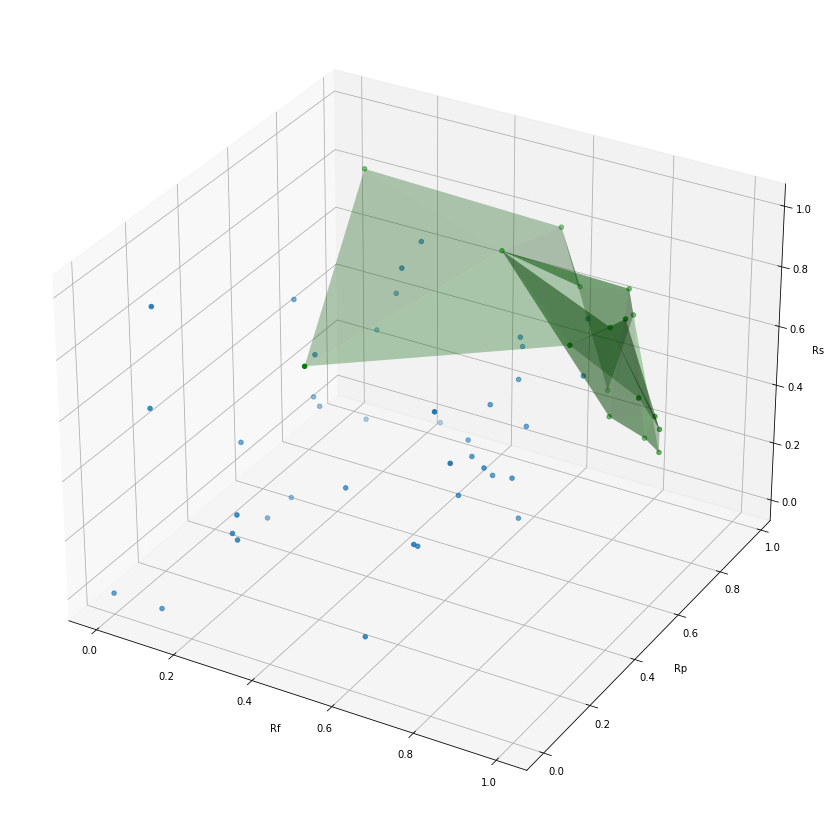}\quad
  \caption{Pareto Fronts for WatDiv$_{full}$}
  \label{fig:1_pareto}
\end{subfigure}\hfil 
\medskip
\begin{subfigure}{0.5\textwidth}
  \includegraphics[trim=0 0 0 0,clip,width=0.8\linewidth]{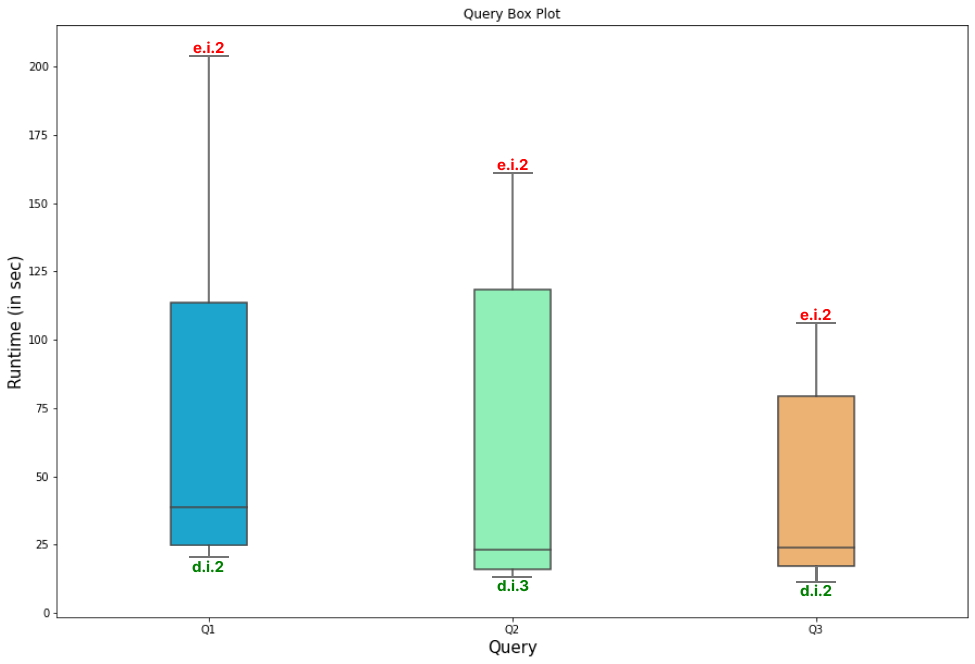}\quad
  \caption{Best and worst-performing configuration for each query}
  \label{fig:2_box}
\end{subfigure}\hfil 
\vspace{-15pt}
\caption{Pareto Fronts, and queries best-worst configuration examples.}
\label{fig:pareto_and_boxplot}

\centering 
\begin{subfigure}{0.35\textwidth}
  \includegraphics[width=0.8\textwidth]{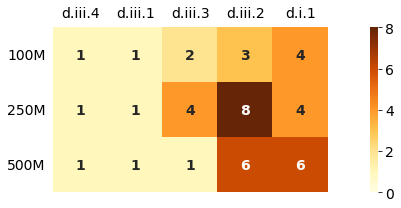}\quad
  \caption{Coherence}
  \label{fig:1_coh}
\end{subfigure}\hfil 
\medskip
\begin{subfigure}{0.65\textwidth}
\centering
\includegraphics[width=0.7\textwidth]{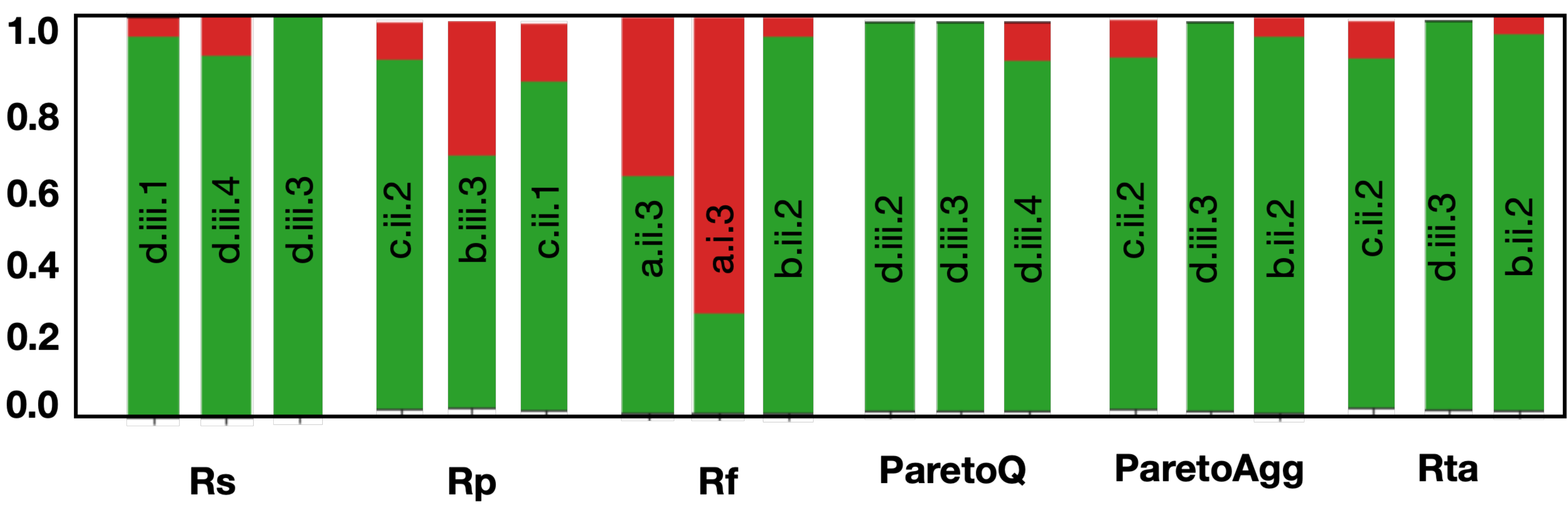}\quad
  \caption{Conformance}
  \label{fig:2_conf}
\end{subfigure}\hfil 
\vspace{-15pt}
\caption{Heatmap shows the coherence of the $R_s$ criterion (Top-5 configurations) scaling from 100M to larger dataset scales. The stacked plot shows the Conformance of the top-3 ranked configurations.}
\label{fig:confStacked_coherHatmap}

\end{figure}

To fulfill \ref{req:viz}, \pap decouples data analytics from visualizations. Meaning that the user can specify his/her visualizations of interest with the performance data. Nevertheless, \pap yet provides several interactive and extensible default visualizations that help practitioners rationalize the performance results and final prescriptions. In addition, visualizations are simple and intuitive for understanding several Bench-Ranking definitions, equations, and evaluation metrics. 
For instance, Figure~\ref{fig:sd} (a-c) shows three samples of SD-ranking criteria plots. In particular, they show how many times a specific dimension's (e.g., schema in Figure~\ref{fig:sd} (a) ) alternatives/options (ST, VP, PT,..etc) achieve the highest or the lowest ranking scores. Figure~\ref{fig:triangles} shows the SD ranking criteria w.r.t a simple geometrical representation (detailed in the following sections) that depicts the triangle subsumed by each dimension's ranking criterion (i.e., Rs, Rp, and Rf). The triangle sides present the trade-offs ranking dimensions and show that the SD-ranking criteria may only optimize towards a single dimension at a time.
The MD-ranking criteria, i.e., Pareto$_{Agg}$~\footnote{Pareto$_{Q}$ cannot be visualized, i.e., as it uses more than \textit{three} dimensions, one for each query of the workload~\cite{ragabieee21}.} results are depicted using a \textit{3D} plot in Figure~\ref{fig:pareto_and_boxplot} (a). Pareto fronts are depicted by the green shaded area of the three experimental dimensions of the Bench-Ranking (for WatDiv 500M triples dataset~\footnote{Due to space limits, we keep other Pareto figures on the \pap GitHub page}). Each point of this figure represents a \textit{solution} of rank scores (i.e., a configuration in our case).

\pap visualizations allow explaining the conformance and coherence results using simple plots. For instance, Figure~\ref{fig:confStacked_coherHatmap} (a) shows the coherence of the top-5 ranked configurations of the Rs criterion in the 100M dataset while scaling to the larger datasets, i.e., 250M and 500M.
\pap explains the conformance of the Bench-Ranking criteria by visualizing the conformance of the top-3 ranked configurations (or any arbitrary number of configurations) with the actual query rankings (Table~\ref{tab:queryruntimes_rankings}). The green color represents the level of conformance, and the red depicts a configuration is performing worse than the $h$ worst rankings. Thus, this may explain why $\mathcal{R}_p$ and $\mathcal{R}_f$ criteria have low conformance results in Table~\ref{tab:results_watdiv}, while the other criteria have relatively higher conformance values.  

Practitioners can also use \pap visualizations for \textit{fine-grained} ranking details. For instance, showing the best and worst configurations for each query (as shown in Figure~\ref{fig:pareto_and_boxplot} (b) for example of \textit{three} queries of the WatDiv workload). Such detailed visualizations could help the user rationalize the final prescriptions of \pap.

\subsection{\pap Flexibility \& Extensibility} \label{subsec:flex_extend}

\begin{wrapfigure}[15]{r}{.45\textwidth}
\vspace{-20pt}
\begin{minipage}{0.45\textwidth}
\begin{lstlisting}[language=yaml,frame=single, caption={Configurations yaml file for different experiments in \pap.},label={lst:full_mini_confs},captionpos=b, basicstyle=\scriptsize]
#(1)Full-WatDiv Configurations
dimensions:
  schemas: ["st", "vp", "pt", "extvp", "wpt"]
  partition: ["horizontal","subject","predicate"]
  storage: ["Avro", "CSV", "ORC", "Parquet"]
query: 20
#(2) Mini-WatDiv Configurations
dimensions:
  schemas: ["st", "vp", "pt"]
  partition: ["horizontal", "subject"]
  storage: ["csv", "orc", "parquet"]
query: 10
#(3)WatDiv without Partitioning (i.e.,Centralized)
dimensions:
  schemas: ["st", "vp", "pt", "extvp", "wpt"]
  partition: null
  storage: ["Avro", "CSV", "ORC", "Parquet"]
query: 20
\end{lstlisting}
\end{minipage}
\end{wrapfigure}

\noindent \textbf{Adding/Removing Configurations/Queries.} 
To show an example of the extensibility of \pap, we implement the Bench-Ranking criteria over a subset of the configurations and subset of the WatDiv benchmark tasks (i.e., queries); we call it \textit{WatDiv$_{mini}$}. In particular, we run \pap Bench-Ranking with WatDiv excluding two schemas (i.e., schema advancements: \textit{ExtVP}, and \textit{WPT}), one partitioning technique (i.e., Predicate-based), and one storage format (i.e., Avro). Then, we include all the configurations back to test the extensibility with the WatDiv experiments (see the left part of Table~\ref{tab:results_watdiv}). The configurations' exclusion and inclusion is specified easily from the \textit{YAML} configuration file (as shown in Listing~\ref{lst:full_mini_confs} lines 7-12), i.e., \pap considers only the specified configurations and rank according to them. 

\begin{wraptable}[7]{r}{0.45\textwidth}
\scriptsize
\vspace{-10pt}
\begin{tabular}{l|c|c|c|c|c}
\hline
\textbf{Schema} &
  \textbf{Full conf. Space} &
  \textbf{PBP} &
  \textbf{!PBP} &
  \textbf{CSV} &
  \textbf{!CSV} \\ \hline
\rowcolor[HTML]{B6D7A8} 
\cellcolor[HTML]{C0C0C0}\textbf{Extvp} &
  0.82 &
  0.96 &
  0.75 &
  0.78 &
  0.83 \\ \hline
\rowcolor[HTML]{D9EAD3} 
\cellcolor[HTML]{C0C0C0}\textbf{PT} &
  0.60 &
  \cellcolor[HTML]{FCE5CD}0.32 &
  \cellcolor[HTML]{B6D7A8}0.74 &
  0.69 &
  0.57 \\ \hline
\rowcolor[HTML]{FFF2CC} 
\cellcolor[HTML]{C0C0C0}\textbf{VP} &
  0.55 &
  \cellcolor[HTML]{D9EAD3}0.71 &
  0.46 &
  \cellcolor[HTML]{B6D7A8}0.74 &
  0.48 \\ \hline
\cellcolor[HTML]{C0C0C0}\textbf{ST} &
  \cellcolor[HTML]{FCE5CD}0.32 &
  \cellcolor[HTML]{FFF2CC}0.52 &
  \cellcolor[HTML]{F4CCCC}0.23 &
  \cellcolor[HTML]{F4CCCC}0.20 &
  \cellcolor[HTML]{FCE5CD}0.37 \\ \hline
\rowcolor[HTML]{F4CCCC} 
\cellcolor[HTML]{C0C0C0}\textbf{WPT} &
  0.21 &
  0.00 &
  \cellcolor[HTML]{FCE5CD}0.31 &
  0.09 &
  0.25 \\ \hline
\end{tabular}
\caption{Schemas global ranking across various configurations.}
\label{tab:global_ranking}
\end{wraptable}

The right side of Table~\ref{tab:results_watdiv} shows the top-ranked \textit{three} configurations according to the specified configurations. Intuitively, results differ according to the available ranked configuration space. For instance, with the inclusion of the ExtVP schema (i.e., '$d$'), it dominates instead of the PT schema (i.e., '$c$') in \textit{WatDiv$_{mini}$} for ranking by schema (Rs) criterion. In \textit{WatDiv$_{mini}$}, excluding the Predicate partitioning ('$iii$'), the \textit{subject-based} partitioning ('$ii$') completely dominates (one exception) in the $\mathcal{R}_p$ criterion across the different dataset sizes. Including it back, the predicate-based partitioning ('$iii$') significantly competes with the \textit{subject-based} partitioning technique in most of the ranking criteria, i.e., $\mathcal{R}_p$, $\mathcal{R}_s$, Pareto$_{Agg/Q}$, and $\mathcal{R}_{ta}$.

With such flexibility, \pap also provides several dynamic views on the ranking criteria. For example, Table~\ref{tab:global_ranking} shows the SD ranking of the schema dimension with changing the configuration space. Particularly, it shows how the \textit{global} ranking of each relational schema (or any other specified dimension) could change by including/excluding configurations of the other dimensions. The table shows that the order of the global schema ranks changes by including 
all configurations ("Full Conf. Space") than including/excluding the predicate partitioning or CSV format, i.e., "PBP/!PBP", "CSV/!CSV", respectively. For instance, the PT schema global ranking is interestingly oscillating with those changes in the available configurations.  

\begin{table}[!htb]
\scriptsize

\begin{minipage}{.5\linewidth}
  \centering
\begin{tabular}{|c|c|c|c|c|}
\hline
\rowcolor[HTML]{C0C0C0} 
\multicolumn{1}{|l|}{\cellcolor[HTML]{C0C0C0}$D_i$} &
  \multicolumn{1}{l|}{\cellcolor[HTML]{C0C0C0}$\mathcal{R}_s$} &
  \multicolumn{1}{l|}{\cellcolor[HTML]{C0C0C0}$\mathcal{R}_f$} &
  \multicolumn{1}{l|}{\cellcolor[HTML]{C0C0C0}Pareto$_{Q}$} &
  \multicolumn{1}{l|}{\cellcolor[HTML]{C0C0C0}Pareto$_{Agg}$} \\ \hline
\cellcolor[HTML]{EFEFEF}                       & d.i.1                         & a.i.3 & c.i.2 & c.i.2 \\
\cellcolor[HTML]{EFEFEF}                       & c.i.2                         & b.i.2 & d.i.2 & b.i.2 \\
\multirow{-3}{*}{\cellcolor[HTML]{EFEFEF}100M} & d.i.2                         & e.i.4 & d.i.4 & d.i.1 \\ \hline
\cellcolor[HTML]{EFEFEF}                       & c.i.1                         & a.i.3 & d.i.2 & d.i.2 \\
\cellcolor[HTML]{EFEFEF}                       & d.i.2                         & e.i.4 & c.i.4 & c.i.1 \\
\multirow{-3}{*}{\cellcolor[HTML]{EFEFEF}250M} & c.i.3                         & d.i.2 & c.i.2 & e.i.4 \\ \hline
\cellcolor[HTML]{EFEFEF}                       & d.i.2                         & a.i.3 & d.i.2 & d.i.2 \\
\cellcolor[HTML]{EFEFEF}                       & \cellcolor[HTML]{FFFFFF}c.i.3 & d.i.2 & c.i.3 & a.i.3 \\
\multirow{-3}{*}{\cellcolor[HTML]{EFEFEF}500M} & \cellcolor[HTML]{FFFFFF}c.i.1 & e.i.4 & c.i.4 & c.i.3 \\ \hline
\end{tabular}
  \caption{Best-performing configurations, \textbf{excluding the partitioning} dimension.}
  \label{tab:watdiv_removing_part_topconfs}
    \end{minipage}%
    \begin{minipage}{.5\linewidth}
      \centering
        
\begin{tabular}{|c|l|c|c|c|c|}
\hline
\rowcolor[HTML]{C0C0C0} 
\multicolumn{1}{|l|}{\cellcolor[HTML]{C0C0C0}\textbf{Metric}} &
  \textbf{D} &
  \multicolumn{1}{l|}{\cellcolor[HTML]{C0C0C0}\textbf{$\mathcal{R}_s$}} &
  \multicolumn{1}{l|}{\cellcolor[HTML]{C0C0C0}\textbf{$\mathcal{R}_f$}} &
  \multicolumn{1}{l|}{\cellcolor[HTML]{C0C0C0}\textbf{Pareto$_{Q}$}} &
  \multicolumn{1}{l|}{\cellcolor[HTML]{C0C0C0}\textbf{Pareto$_{Agg}$}} \\ \hline
\cellcolor[HTML]{EFEFEF}                              & 100M      & 97.50\%  & 67.50\% & 95.00\%  & 92.50\% \\ \cline{2-6} 
\cellcolor[HTML]{EFEFEF}                              & 250M      & 97.50\%  & 30.00\% & 100.00\% & 97.50\% \\ \cline{2-6} 
\multirow{-3}{*}{\cellcolor[HTML]{EFEFEF}Conform.} & 500M      & 100.00\% & 52.50\% & 100.00\% & 52.50\% \\ \hline
\cellcolor[HTML]{EFEFEF}                              & 100M-250M & 0.11     & 0.17    & 0.11     & 0.19    \\ \cline{2-6} 
\cellcolor[HTML]{EFEFEF}                              & 100M-500M & 0.28     & 0.16    & 0.30     & 0.18    \\ \cline{2-6} 
\multirow{-3}{*}{\cellcolor[HTML]{EFEFEF}Cohere.}   & 250M-500M & 0.17     & 0.08    & 0.15     & 0.09    \\ \hline
\end{tabular}
\caption{Criteria evaluation (conform.ance, and coher.ence), \textbf{excluding partitioning}.}
\label{tab:watdiv_removing_part_metrics}
\end{minipage} 
    

\end{table}

\begin{wrapfigure}[10]{r}{0.35\textwidth}
\vspace{-10pt}
    \centering
      \includegraphics[width=0.75\linewidth]{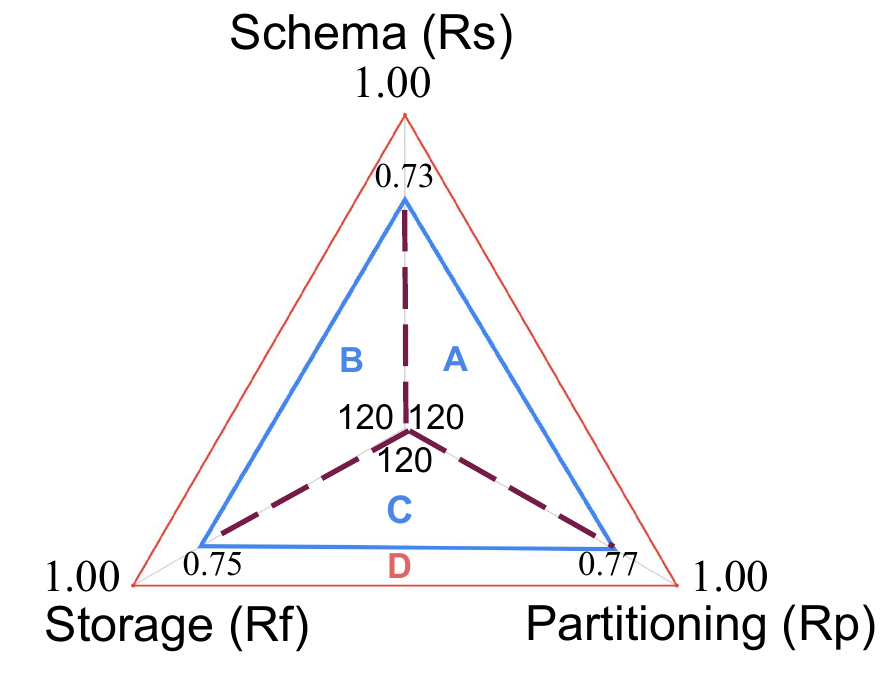}
      \caption{Triangle Area criterion.}
      \label{fig:RaTriangle}
\end{wrapfigure} 

\vspace{3pt}
\noindent \textbf{Adding/Removing Full Experimental Dimension.} \pap'flexibility extends to the experimental dimensions, i.e., it is possible to add/remove dimensions easily. For instance, we can fully exclude the partitioning dimension in case experiments are executed on a single machine (see Listing~\ref{lst:full_mini_confs} lines 13-18). In~\cite{ragab2019benchmarking}, we run experiments on SparkSQL with different relational schemas and storage backends yet without data partitioning. Table~\ref{tab:watdiv_removing_part_topconfs} shows the best-performing (top-3) configurations in WatDiv experiments when excluding the partitioning dimension. Table~\ref{tab:watdiv_removing_part_metrics} shows the conformance and coherence metrics' results for the various ranking criteria~\footnote{Notably, the Rp and Rta criteria cannot be calculated when excluding the partitioning dimension.}. 

\vspace{2pt}
\begin{wrapfigure}[17]{r}{.42\textwidth}
\vspace{-20pt}
\begin{minipage}{0.42\textwidth}
\scriptsize
\begin{lstlisting}[language=python, frame=single, caption={Plugin the Triangle-Area as new Ranking criterion.},label={lst:newRanker},captionpos=b, basicstyle=\scriptsize]
from papya import Rank
#SD Ranker (Implementation of Equation (1))
class SD_Ranking(Rank):
    ...
#MD Ranker (Pareto-NSGA2)    
class MD_Ranking(Rank):
    ...    
#Add Triangle_Area as a new ranking criterion.
class RtaCriterion(Rank):
  def calculate_rta(self):
     r_scores = SDRank(...).calculateRank()
     rta_scores = []
     for i in range (len(r_scores)):
       rs = r_scores[0][i]
       rp = r_scores[1][i]
       rf = r_scores[2][i]
       # RTA Formula (Equation 4)
       y = (math.sin(math.radians(120)))/2
       outer_triangle_area = y * (1+1+1)
       rta = y * (rf*rp + rs*rp + rf*rs)
       rta_scores.append(rta)
    return rta_scores
  def plot(self):# plots (Figure 7)
    ...
\end{lstlisting}
\end{minipage}
\end{wrapfigure}
\noindent \textbf{Adding Ranking Criterion.} \pap abstractions enable users to plug-in new ranking criterion besides the already existing ones (i.e., the abstract Rank class, Section~\ref{sec:lib_req_arch}). Let's assume we seek usage of a simple ranking criterion that leverages a \textit{geometric} interpretation of the SD rankings of the three experimental dimensions based on the \textit{triangle area} subsumed by each ranking criterion ($R_s$, $R_p$, and $R_f$).

In Figure~\ref{fig:RaTriangle}, the triangle sides represent the SD-ranking dimensions' rank scores. Thus, this criterion aims to maximize this triangle's area (i.e., the \textit{blue} triangle). The closer to the ideal (outer red triangle), the better it scores. In other words, the bigger the area of this triangle covers, the better the performance of the three ranking dimensions altogether. The \textit{red} triangle represents the case with the maximum/ideal rank score, i.e, $R=1$ for the three dimensions (as, $0 < R <= 1$). Equation (1) defines the blue triangle area; we call it ranking by triangle area ($R_{ta}$). 

   \begin{ceqn} \label{eq:inner_triangle}
    \begin{align}          
        {\displaystyle TriangleArea (R_{ta})=\frac{1}{2} sin(120)*(R_f*R_p+R_s*R_p+R_f*R_s)}
    \end{align}
    \end{ceqn}

The formula (Cf. Equation (4) ) computes the actual triangle area. Simply, it sums up the triangle area of the three triangles $A$, $B$, and $C$ by two of its sides which are the rank scores of each dimension, i.e., $R_s$, $R_p$, or $R_f$ (dashed triangle sides), and the angle between both of them (i.e $120$ in this case). 
For example, the actual area of the \textit{blue} triangle is $R_{ta} =\frac{1}{2}\sin(120) (0.75*0.771+0.73*0.77+0.75*0.73) = 0.73$. 
In addition to the SD and MD ranking criteria classes, Listing~\ref{lst:newRanker} shows how to extend \pap with a new Ranker class, i.e., \textit{RtaCriterion}.

It is worth mentioning that the idea behind $R_{ta}$ is intuitively similar to Pareto$_{Agg}$ because both aim to maximize the rank scores of the three dimensions altogether. However, unlike Pareto$_{Agg}$ that is multi-dimensional, $R_{ta}$ cannot extend to dimensions above three. For simplicity, we used $R_{ta}$ as an exemplar to showcase that \pap abstractions enable extending new ranking algorithms/criteria. Table~\ref{tab:results_watdiv} shows the top-5 best-performing configurations ranked by $R_{ta}$. Results show that Pareto$_{Agg}$ results perfectly conform with $R_{ta}$ top-ranked configurations (especially in the top-3 ranked configurations). Table~\ref{tab:eval_metrics_watDiv_mini_full} also show that $R_{ta}$ criterion scores high conformance ratios across WatDiv benchmark datasets. It also scores high coherence (few disagreements) through the scalability of WatDiv datasets. In both WatDiv experiments (i.e., with mini and full dimensions inclusion), the $R_{ta}$ conformance and coherence values are very close to the Pareto$_{Agg}$ criterion. 

\subsection{Checking Performance Replicability} \label{sec:replicability}

\begin{figure}[t!]
    \centering 
\begin{subfigure}{0.5\textwidth}
  \includegraphics[trim=0 0 0 0,clip,width=0.75\linewidth]{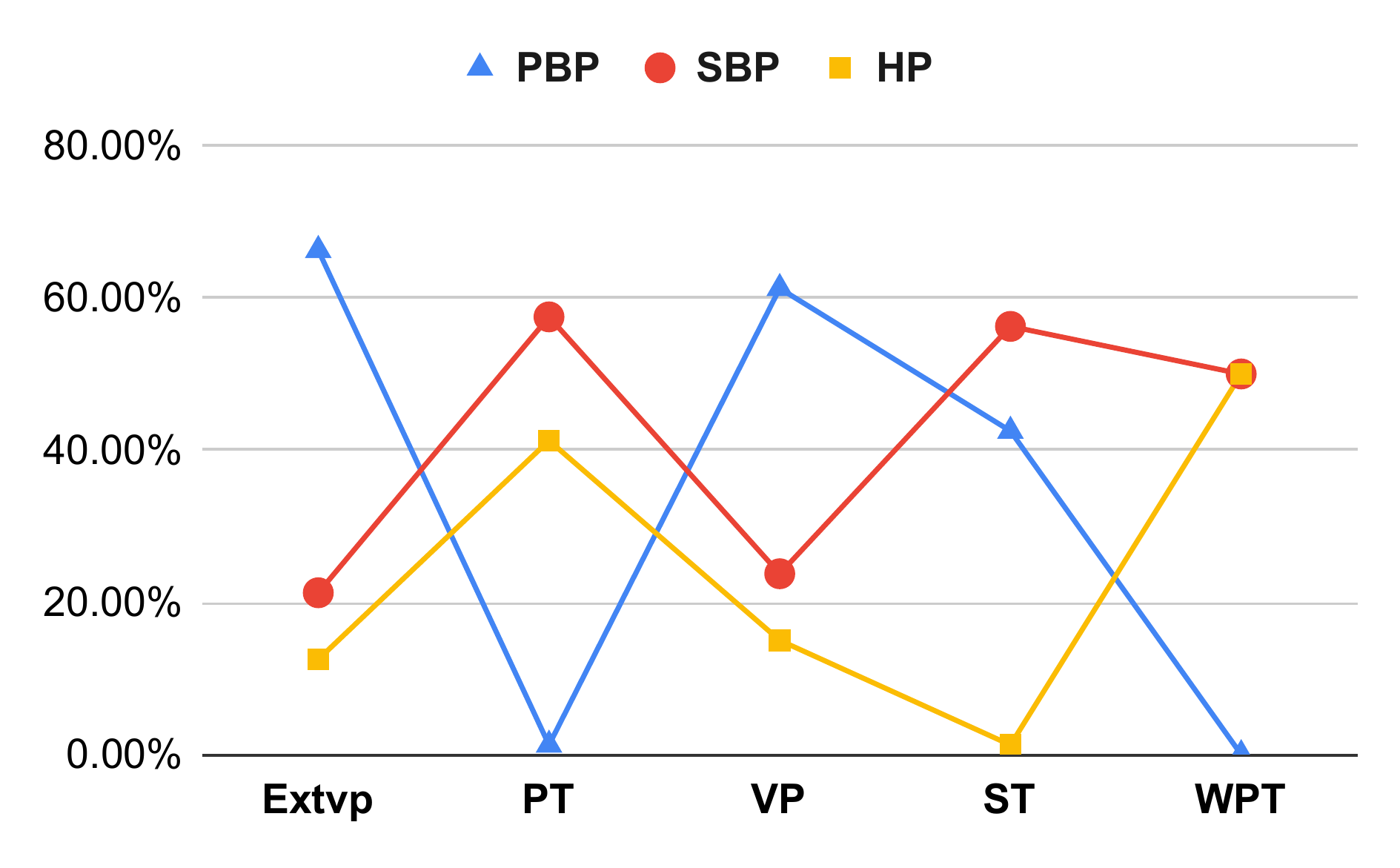}\quad
  \caption{Impact of partitioning on the schema performance}
  \label{fig:1_partimapct}
\end{subfigure}\hfil 
\medskip
\begin{subfigure}{0.5\textwidth}
  \includegraphics[trim=0 0 0 0,clip,width=0.75\linewidth]{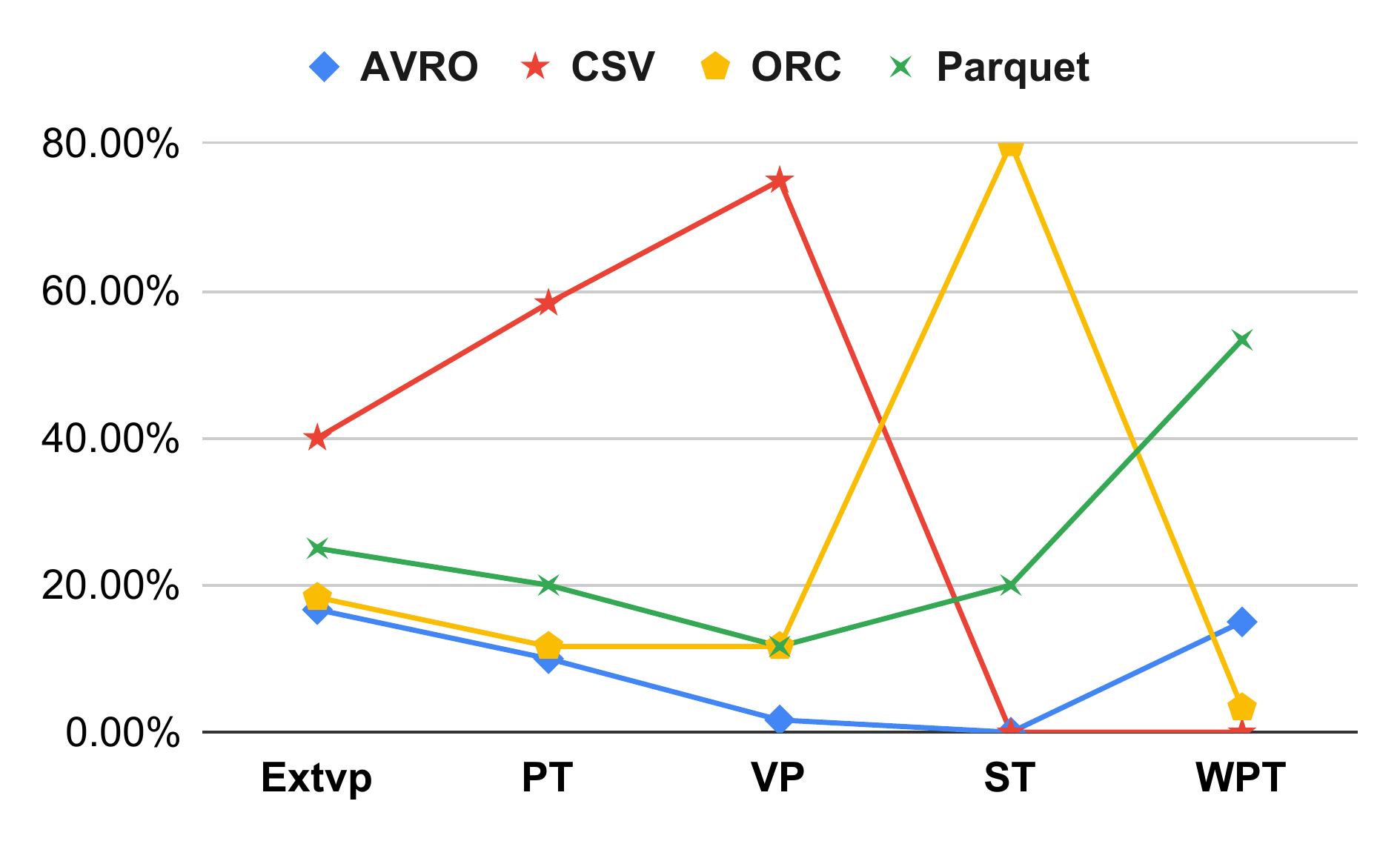}\quad
  \caption{Impact of storage on the schema performance}
  \label{fig:2_storageimpact}
\end{subfigure}\hfil 
\caption{Schema Replicability across changing partitioning or storage formats.}
\label{fig:replicability_schema}
\vspace{-5pt}
\end{figure}

\pap also activates the functionality of checking the BD system's performance replicability when introducing different experimental dimensions. In particular, it enables checking the system's performance with one specific dimension while changing the parameters of the other dimensions. For example, Figures~\ref{fig:replicability_schema} (a) and (b) respectively show the impact of the partitioning and storage on the performance of the schema dimension. The Figures show how the performance of the system with a configuration can significantly change with changing other dimensions. 

\pap can also check the performance replicability by comparing two configurations as discussed in~\cite{ragab2021depth}. For instance, \pap can compare the \textit{schema} optimizations (i.e., \textit{WPT}, and \textit{ExtVP}) w.r.t their \textit{baseline} ones (i.e., \textit{PT}, and \textit{VP}) while introducing different partitioning techniques and various HDFS storage formats that are different from the baseline configurations~\cite{schatzle2016s2rdf,schatzle2014sempala}.

\begin{wraptable}[7]{r}{0.55\textwidth}
\scriptsize
\vspace{-5pt}
\scriptsize
\centering
\setlength\tabcolsep{1.5pt}

\begin{tabular}{|c|l|c|c|c|l|c|c|}
\hline
 & Partitioning          & ExtVP VS.VP & WPT VS. PT  &  & Storage & ExtVP VS. VP & WPT VS. PT  \\ \cline{2-4} \cline{6-8} 
 &
  \cellcolor[HTML]{B6D7A8}V. HDFS &
  \cellcolor[HTML]{B6D7A8} 97.5\% &
  \cellcolor[HTML]{B6D7A8} 54.16\% &
   &
  \cellcolor[HTML]{B6D7A8}Parquet &
  \cellcolor[HTML]{B6D7A8}75.0\% &
  \cellcolor[HTML]{B6D7A8} 38.8\% \\ \cline{2-4} \cline{6-8} 
 & Horizontal      & 67.5\%     & 8.3\% &  & ORC    & 73.33\%     & 18.5\% \\ \cline{2-4} \cline{6-8} 
 & Predicate & 61.4\%     & NA      &  & Avro   & 63.3\%  & 16.6\% \\ \cline{2-4} \cline{6-8} 
\multirow{-5}{*}{\rotatebox[origin=c]{90}{Partitioning}} &
  Subject &
  66.25\% &
  6.9\% &
  \multirow{-5}{*}{\rotatebox[origin=c]{90}{Storage}} &
  CSV &
 93.3 \% &
 16.6 \% \\ \hline
\end{tabular}
\caption{The \textit{replicability} of schema advancements (i.e., WPT, ExtVP) VS. baselines (i.e., PT, VP), WatDiv 500M dataset.}
\label{tab:replicability}
\vspace{-10pt}
\end{wraptable}

Table~\ref{tab:replicability} shows the effect of introducing partitioning techniques (right of the table) and different file formats (left of the table) different from the baseline configurations (i.e., Vanilla HDFS partitioning, and Parquet as storage format). The \textit{trade-offs} effect is clear on the replicability results. Indeed, WPT outperforms PT schema only with $54.16\%$ in the queries using the baseline Vanilla HDFS partitioning technique across all storage formats and only about $39\%$ for the baseline Parquet format across all partitioning techniques. On the other side, we observe significant degradation of WPT schema optimization moving to other configurations with both partitioning and storage dimensions. For instance, WPT outperforms PT only with about~$8\%$ and~$7\%$ using other different partitioning techniques, i.e., Horizontal and Subject, respectively. The same occurs with changing the storage formats different from baseline Parquet. Similarly, ExtVP versus VP schema performance results confirm our observations. \pap enables showing those \textit{trade-offs} of considering alternative storage file formats and partitioning techniques alongside the experiments’ query evaluation.

\vspace{-15pt}

\section{Related Work}\label{sec:relwork}

The Semantic Web community is synergizing efforts pushing technological progress through benchmarking initiatives like the Linked Data Benchmark Council (LDBC\footnote{ldbcouncil.org}), and providing full-fledge benchmarks like \textit{SP$^2$Bench}~\cite{schmidt2009sp}, \textit{WatDiv}~\cite{alucc2014diversified} to name few. 
In the literature, several tools and frameworks seek to reduce the effort required to design and execute reproducible experiments, share their results, and build pipelines for various Semantic Web applications. For instance, the \textit{RSPLab} provides a \textit{test-drive} for stream reasoning engines that can be deployed on the cloud. It enables the design of experiments via a programmatic interface that allows deploying the environment, running experiments, measuring the performance, and visualizing the results. 
In the work in~\cite{acosta2017diefficiency}, the authors tackled the problem of quantifying the continuous behavior of a query engine and presented two novel experimental metrics that  
capture the performance efficiency during an elapsed period rather than a constant instance of time. On another side, authors in~\cite{ayala2018flexible} provide a data-loader that facilitates generating RDF graphs in different logical partitioning schemas and physical partitioning options. However, this tool stops the work of data generation and data loading. That is, this tool leaves the work of deciding the best solutions for the data/knowledge engineers.
 
The mentioned efforts aim at providing the environment that enables the practitioners to develop their experimental pipelines. Nonetheless, none of these efforts provide prescriptive performance analyses in the context of BD problems. Conversely, \pap aims to cover this timely research gap in an easy and extensible approach, facilitating building the experimental solution space, and automating providing prescriptions over it.

\vspace{-15pt}

\section{Conclusion and Roadmap}\label{sec:conclusion}
This paper presents \pap, an extensible library that reduces the efforts needed to analyze the performance of BD systems used for processing large (RDF) graphs. \pap implements the performance analytics methods adopted in~\cite{schatzle2014sempala,schatzle2016s2rdf,ragab2019benchmarking} including an novel approach for prescriptive performance analytics we presented in~\cite{ragabieee21}.

\begin{figure*}[t!]
\vspace{-10pt}
    \centering
     \includegraphics[width=0.45\linewidth]{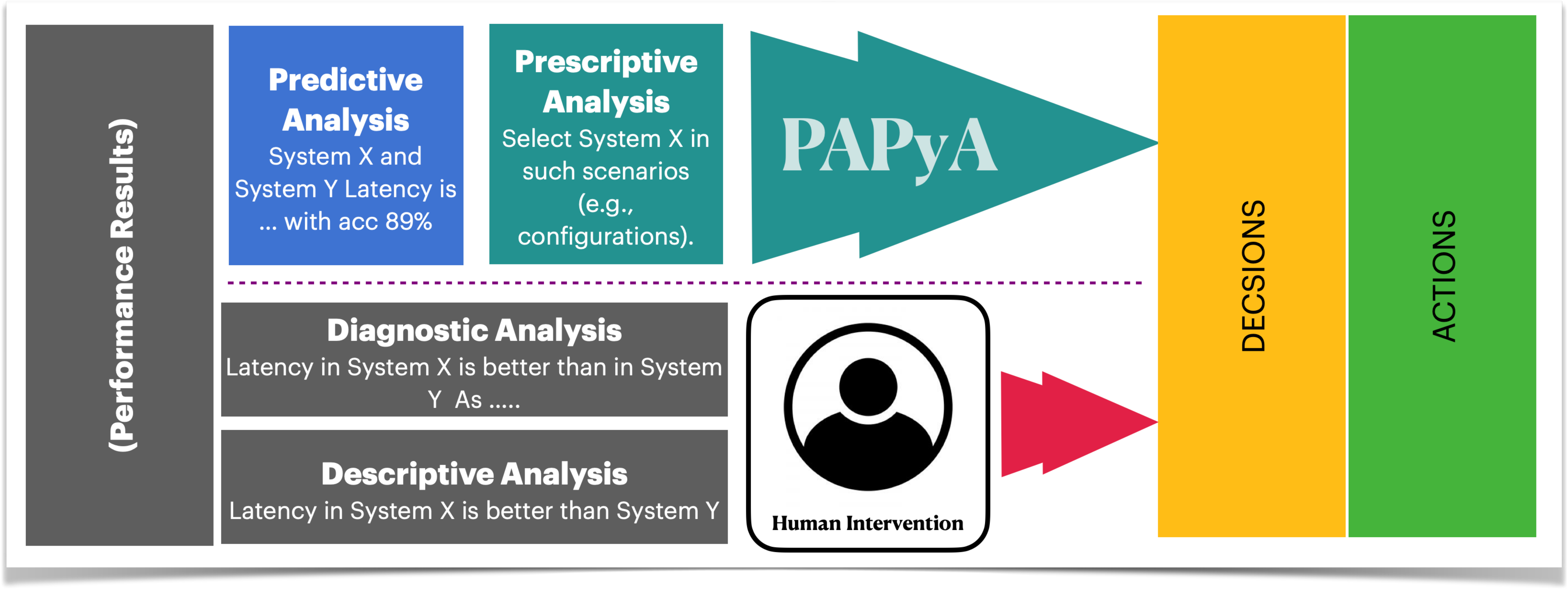}
      \caption{Performance analysis methodology, and how \pap reduces human intervention in BD performance analyses.}
      \label{fig:mygartner}
\end{figure*}

Inspired by Gartner's analysis methodology~\cite{gartner}, Figure~\ref{fig:mygartner} reflects the amount of human intervention required to make a decision with the descriptive and diagnostic analyses of the performance results. Descriptive and diagnostic analytics are limited, and cannot guide practitioners directly to the best-performing configurations in a complex solution space. This is shown in this paper with the lack of performance replicability (shown Section~\ref{sec:replicability}). Indeed, the performance of the BD system is affected by changing the configurations, e.g., oscillating schema performance with changing partitioning, and storage options (Figure~\ref{fig:replicability_schema}).    
On the other side, \pap reduces the amount of work required to interpret performance data. It adopts the Bench-ranking methodology with which practitioners can easily decide the \textit{best-performing} configurations given an experimental solution space with an arbitrary number of dimensions. Although, descriptive discussions are limited, \pap still provides several descriptive analytics and visualizations on the performance to explain the final decisions given by \pap. 
\pap also aims to reduce the engineering work required for building an analytical pipeline for processing large RDF graphs. In particular, \pap prepares, generates, and loads data ready for big relational RDF graph analytics.

\pap is developed considering the ease of use and the flexibility aspects allowing extending the library with an additional arbitrary number of experimental dimensions to the solution space. Moreover, \pap provides abstractions on the level of ranking criteria, meaning that the user can use his/her ranking functions for ranking the solution space. Seeking availability, we provide \pap as an open-source library under MIT license and published at a persistent URI. \pap's GitHub repository includes tutorials and documentation on how to use the library. 

As a maintenance plan, \pap's roadmap includes:
\begin{enumerate}[leftmargin=*]

\item covers the phase of query evaluation into \pap pipeline. In particular, we plan to provide native support of SPARQL via incorporating native triple stores for query evaluation. 

\item incorporate SPARQL into SQL translation using advancements of \textit{R2RML} mapping tools (e.g., \textit{OnTop})~\cite{belcao2021chimera}.

\item wraps other \textit{SQL-on-Hadoop} executors to \pap; thus, the performance of the engines could also be compared.

\item use orchestration tools (such as \textit{Apache Airflow}) to monitor the \pap pipelines.

\item integrate \pap with other tools like~\textit{gmark}~\cite{bagan2016gmark} that generates graphs and workloads, making them ready for distributed relational setups.
\end{enumerate}

\noindent\small\textbf{Acknowledgments.}
We acknowledges support from the European Social Fund via IT Academy programme and the European Regional Development Funds via the Mobilitas Plus programme (grant MOBTT75).

\vspace{-15pt}

\nocite{*}
\bibliographystyle{ios1}           
\bibliography{bibliography}        

\end{document}